\documentclass[%
 reprint,onecolumn,
superscriptaddress, 
12pt, 
nofootinbib,
 amsmath,amssymb,
floatfix,
]{revtex4-2}

\usepackage{graphicx}
\usepackage{subcaption} 
\usepackage{dcolumn}
\usepackage{bm}

\usepackage{cancel}
\usepackage{amsfonts}
\usepackage{xcolor}
\usepackage{tcolorbox}

\newcommand{\beqa}{\begin{eqnarray}}
\newcommand{\eeqa}{\end{eqnarray}}
\newcommand{\nn}{\nonumber}

\begin{document}

\title{Hamiltonian Active Particles in Incompressible Fluid Membranes}

\author{Sneha Krishnan}
\affiliation{Birla Institute of Technology and Science, Pilani, Hyderabad Campus, Telangana 500078, India}
\author{Rickmoy Samanta}
\affiliation{Birla Institute of Technology and Science, Pilani, Hyderabad Campus, Telangana 500078, India}
\affiliation{Indian Institute of Technology Kharagpur, West Bengal 721302, India}

\date{\today}
\begin{abstract}
Active proteins and membrane-bound motors exert force dipole flows along fluid interfaces and lipid bilayers. We develop a Hamiltonian framework for the interactions of pusher and puller dipoles embedded in an \emph{incompressible} two-dimensional membrane supported by a shallow viscous subphase. Beginning from the Brinkman-regularized Stokes equations of the membrane-subphase system, we construct the near-field and far-field dipolar velocity fields and associated stream functions. For two quenched (fixed orientation) dipoles, we obtain exact analytic solutions in both the near- and far-field regimes. Although generic dipoles reorient under local membrane vorticity, we show that the far-field dipolar flow is vorticity-free; force-free motors therefore retain fixed orientations and obey a position-based Hamiltonian dynamics in which the positions of $N$ dipoles evolve via an effective Hamiltonian built from the dipolar stream function. In the near field, where the flow possesses finite vorticity, a Hamiltonian formulation is recovered in the quenched-orientation limit. For identical dipoles, the far-field Hamiltonian produces rapid clustering from random initial conditions, whereas the near-field Hamiltonian suppresses collapse leading to non-aggregating configurations. The phase portraits reveal that the many-body collapse observed in the time evolution of the far field Hamiltonian arises from angular transport across phase space enabled by the screened hydrodynamic interaction. Our work thus provides a concrete realization of a position-based Hamiltonian formulation for active particles in incompressible fluid membranes and shows that hydrodynamic screening can reorganize both the dynamical phase space and the collective organization of active dipoles.
\end{abstract}
\maketitle

\section{Introduction}
\label{sec:intro}
Active proteins, enzymes, and membrane-bound motors continually inject stress into biological interfaces, driving flows along fluid membranes and lipid bilayers.  
At the micron scale, inertia is negligible, and these inclusions generate flow fields whose leading contribution is a force dipole (stresslet) ~\cite{purcell,lauga2009}.  
Such dipoles can be either extensile ``pushers'' or contractile ``pullers''. They play central roles in pattern formation, instabilities, and collective behaviour across active fluids~\cite{Hatwalne2004,Vicsek1995,Marchetti2013}.  
Membrane-bound proteins and enzymatic assemblies act as force dipoles to leading order~\cite{mik}, and experiments have demonstrated motor–mediated clustering and cooperative transport~\cite{aggexp1,aggexp2}.  When confined to a two-dimensional interface, these dipoles interact very differently from their bulk counterparts, primarily due to planar geometry and momentum exchange with the surrounding fluid.

Supported membranes provide a candidate setup in which these effects can be investigated explicitly.  Classical work by Saffman, Delbrück, and subsequent authors~\cite{saff1,saff2,hughes,evans,lg96,staj,fischer,OppenheimerDiamant2009,OppenheimerDiamant2010,OppenheimerDiamant2011} showed that a viscous membrane coupled to a shallow subphase obeys a Brinkman-regularized Stokes equation.  
The corresponding Green’s function exhibits a crossover set by a hydrodynamic screening length: an unscreened logarithmic regime at short distances and a screened regime at large distances.  At the level of dipolar interactions, this produces a transition from $1/r$ to $1/r^3$ flows, accompanied by a change in the overall structure of the velocity field. Although recent studies of swimmers and active inclusions in thin films and membranes~\cite{leoni2010,manikantan2020,bagaria2022,jain2023} have emphasized the richness of these interactions, a systematic understanding of how this crossover impacts the dynamical structure of dipole interactions, particularly the  role of fluid vorticity and the emergence of reduced Hamiltonian descriptions, remains to be established.

In this work, we develop an analytic framework for force-dipole interactions in an \emph{incompressible} supported membrane and show that the screened hydrodynamic interaction leads to two qualitatively distinct dynamical regimes.  
Starting from the Brinkman-regularized membrane Stokes equation, we derive the real-space Green’s tensor and construct the associated dipolar velocity fields and stream functions in both regimes.  
In the near field, the dipolar flow is purely radial at leading order and carries finite vorticity. As a consequence, the relative orientation of a dipole pair is conserved only in the quenched (fixed orientation) limit, and the two-body dynamics becomes exactly solvable, with the squared separation evolving linearly in time.  
In the far field, by contrast, the flow is screened and vorticity-free, and contains both radial and azimuthal components. The resulting two-body problem remains exactly integrable but becomes intrinsically two-dimensional, with coupled radial and angular dynamics governed by a conservation law.

These differences have direct implications for the existence and structure of Hamiltonian formulations.  
For incompressible membranes, the positional dynamics can be written in Hamiltonian form when dipole orientations are fixed. In the far field, the absence of vorticity makes the dipoles \textit{naturally} quenched, yielding an exact position-based Hamiltonian.  
In the near field, finite vorticity generically induces orientational dynamics, so a position-only Hamiltonian arises only in the enforced quenched-orientation limit. At the two-body level, the near-field Hamiltonian is effectively one-dimensional, while the far-field Hamiltonian produces genuine radial–angular coupling.

Exploiting these Hamiltonian structures, we compare the collective behaviour generated by the near- and far-field interactions.  
 The screened far-field interaction drives aggregation, with both pushers and pullers forming compact clusters. In contrast, the unscreened near-field interaction suppresses long-lived aggregation, leading instead to extended configurations with increasing mean separation. A close examination of the phase portraits of the near- and far-field Hamiltonians also reveals the underlying mechanism. The resulting many-body collapse observed in the time evolution of the far field Hamiltonian arises from angular transport across
phase space enabled by the screened hydrodynamic interaction. Taken together, our results show that subphase-induced hydrodynamic screening in  membranes affects the dynamical structure of dipole interactions.  
The system provides a minimal setup in which explicit Hamiltonian descriptions of active particles can be constructed and analyzed, with nontrivial collective behaviour.  
This framework offers a starting point for extensions to more complex membrane rheologies ~\cite{bagaria2022,jain2023,Shoham2023,Hosaka2023,skumrs2025}, including chiral or curved membranes, and suggests experimentally testable consequences in artificial setups where membrane viscosity and screening length can be systematically tuned.

\section{Incompressible supported membrane}
\label{sec:incompressible}
\begin{figure}[t]
    \centering
    \includegraphics[width=0.9\linewidth]{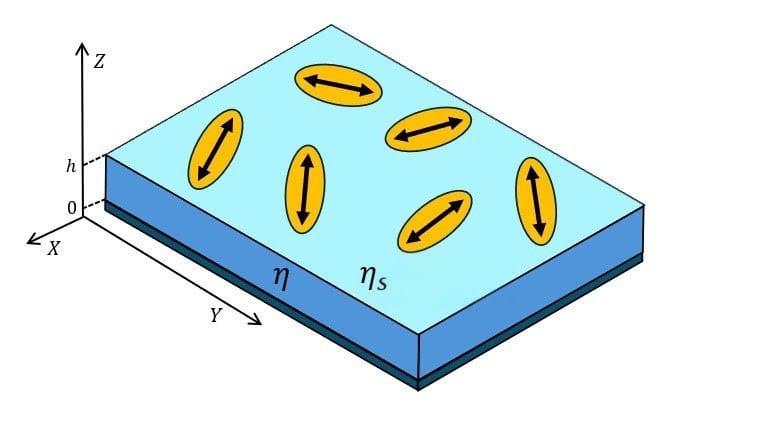}
    \caption{Schematic of active force-dipole motors confined to a supported fluid membrane.  
    Orange ellipses denote motors with orientations (double arrows).  
    The light blue layer is an incompressible two-dimensional membrane of shear viscosity $\eta_s$, supported by a viscous subphase of viscosity $\eta$ and thickness $h$ above a rigid substrate.}
    \label{fig:schematic}
\end{figure}

We consider a two-dimensional viscous membrane of shear viscosity $\eta_s$ lying in the plane $z=h$, supported by a Newtonian shallow subphase of viscosity $\eta$ and thickness $h$ above a rigid substrate (Fig.~\ref{fig:schematic}).  
Throughout the main text we restrict to the incompressible limit: the in-plane velocity field $\mathbf v(\mathbf r)$ satisfies $\nabla\!\cdot\!\mathbf v=0$, there is no dilatational viscosity.  Eliminating the three-dimensional subphase flow under a lubrication approximation yields an effective Brinkman friction acting on the membrane~\cite{staj,Hosaka2023}, so that the steady Stokes balance equation reads
\begin{equation}
\eta_s\nabla^2\mathbf v -\nabla p - \zeta_\parallel \mathbf v + \mathbf F = \mathbf 0,
\qquad
\nabla\!\cdot\!\mathbf v = 0,
\label{eq:membrane_stokes}
\end{equation}
where $p$ is the membrane pressure, $\mathbf F(\mathbf r)$ is any in-plane body force density, and $\zeta_\parallel$ is a phenomenological friction coefficient that encodes momentum leakage into the subphase.  For a shallow film one typically expects
\begin{equation}
\zeta_\parallel \sim \frac{\eta}{h},\nn
\end{equation}
so that the hydrodynamic screening length
\begin{equation}
\kappa^{-1} = \sqrt{\frac{\eta_s}{\zeta_\parallel}}
            = \sqrt{\frac{\eta_s h}{\eta}}\nn
\end{equation}
sets the crossover between ``membrane-dominated'' and ``subphase-dominated'' flow.  Because Eq.~\eqref{eq:membrane_stokes} is linear and translationally invariant in the membrane plane, the velocity response to a localized in-plane force $\mathbf F(\mathbf r) = \mathbf f\,\delta(\mathbf r)$ can be written in terms of a Green's tensor $G_{ij}(\mathbf r)$,
\begin{equation}
v_i(\mathbf r) = G_{ij}(\mathbf r) f_j.\nn
\end{equation}
In Fourier space, membrane incompressibility can be utilized to obtain
\begin{equation}
\tilde G_{ij}(\mathbf q)
 = \frac{1}{\eta_s(q^2+\kappa^2)}
   \big(\delta_{ij} - \hat q_i \hat q_j\big),
\label{eq:Gq}
\end{equation}
where $\hat{\mathbf q}=\mathbf q/q$.  
The real-space Green's tensor is obtained from the inverse transform
\begin{equation}
G_{ij}(\mathbf r)
  = \int\!\frac{d^2q}{(2\pi)^2}\,
    e^{i\mathbf q\cdot\mathbf r}\,
    \tilde G_{ij}(\mathbf q).\nn
\end{equation}
Rotational invariance implies the decomposition
\begin{equation}
G_{ij}(\mathbf r)
= \frac{1}{2\pi\eta_s}
\Big[ A(r)\,\delta_{ij}
    + B(r)\,\hat r_i \hat r_j \Big],
\label{eq:G_realspace}
\end{equation}
with $r=|\mathbf r|$ and $\hat{\mathbf r}=\mathbf r/r$.  
Explicit integration yields
\begin{align}
A(r) &= K_0(\kappa r) + \frac{K_1(\kappa r)}{\kappa r}
       - \frac{1}{\kappa^2 r^2}, \label{eq:A_full} \\
B(r) &= -K_0(\kappa r) - \frac{2K_1(\kappa r)}{\kappa r}
       + \frac{2}{\kappa^2 r^2}, \label{eq:B_full}
\end{align}
where $K_n$ are modified Bessel functions of the second kind.  
Equations~\eqref{eq:G_realspace}-\eqref{eq:B_full} constitute the two-dimensional Stokeslet for an incompressible supported membrane~\cite{staj,fischer}.  
\section{Dipolar flows: near and far zones}
\label{sec:dipolar}
\begin{figure}[t]
\centering
\includegraphics[width=0.9\linewidth]{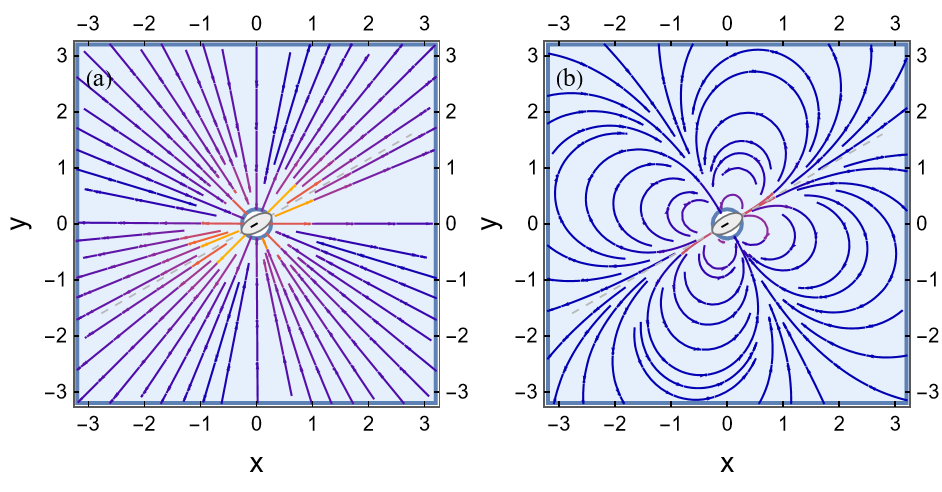}
\caption{
Flow of a single \emph{pusher} dipole in an incompressible supported membrane. 
(a) Near field ($\kappa r \ll 1$): radial $v\sim 1/r$ flow with finite vorticity. 
(b) Far field ($\kappa r \gg 1$): screened $v\sim 1/r^{3}$ flow with both radial and azimuthal components and vanishing vorticity ($r>0$). 
The dashed line indicates the dipole axis.
}
\label{fig:dipole_near_far}
\end{figure}
\begin{figure}[t]
\centering
\includegraphics[width=0.9\linewidth]{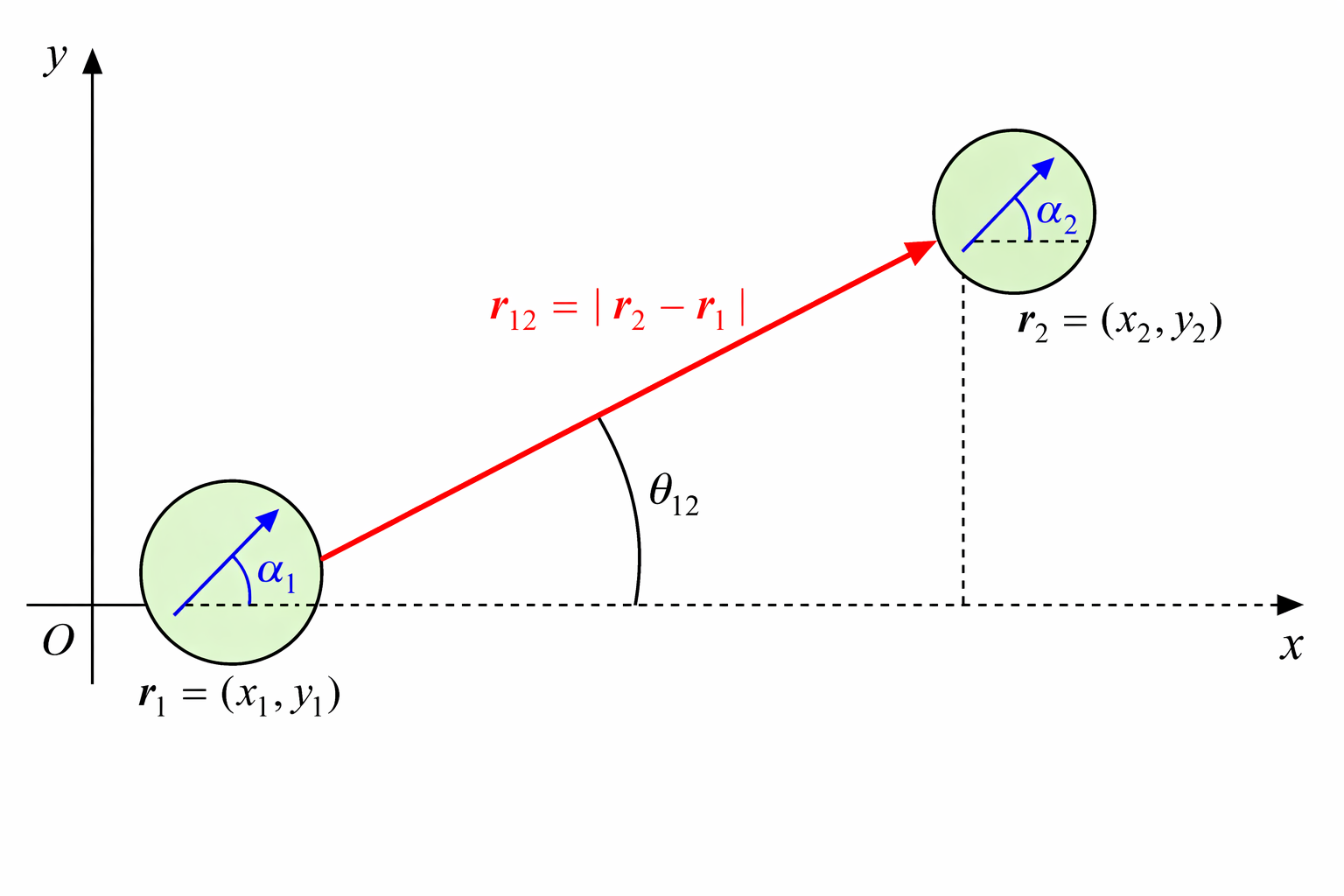}
\caption{Two-body geometry for quenched dipole interactions in a two-dimensional membrane. Dipoles at $\mathbf r_1=(x_1,y_1)$ and $\mathbf r_2=(x_2,y_2)$ have fixed orientations $\alpha_1$ and $\alpha_2$ (blue arrows). 
The separation vector $\mathbf r_{12}=\mathbf r_2-\mathbf r_1$ (red) has magnitude $r_{12}$ and polar angle $\theta_{12}$ measured from the $x$-axis. 
These variables define the angular structure entering the Hamiltonian and equations of motion.}
\label{fig:dipole_schematic}
\end{figure}
Force-free active inclusions such as membrane-anchored motors are naturally modelled as force dipoles (stresslets).  
A point stresslet of strength $\sigma$ and orientation $\hat{\mathbf d}$ located at the origin can be represented as the derivative of the Stokeslet along the dipole axis,
\begin{equation}
v_i(\mathbf r)
 = \sigma\,\hat d_k\,\partial_k G_{ij}(\mathbf r)\,\hat d_j,
\label{eq:dipolar_def}
\end{equation}
with $\hat{\mathbf d}=(\cos\alpha,\sin\alpha)$. Here $\partial_{k}$ denotes differentiation with respect to the source coordinate. The sign of $\sigma$ sets the dipole type: $\sigma>0$ corresponds to a pusher (extensile) and $\sigma<0$ to a puller (contractile). 
\subsection{Near field: unscreened hydrodynamics}
For separations much smaller than the screening length, $\kappa r\ll1$, the Bessel functions can be
expanded as follows
\begin{align}
A(r) &\simeq \frac14\Big[-1-2\gamma
            -2\ln\!\Big(\frac{\kappa r}{2}\Big)\Big], \\
B(r) &\simeq \frac12,\nn
\end{align}
where $\gamma$ is Euler's constant.  
Substituting these into Eq.~\eqref{eq:dipolar_def} and differentiating yields a purely radial near-field flow for a dipole placed at the origin,
\begin{equation}
\mathbf v_{\rm near}(\mathbf r)
 = \frac{\sigma}{4\pi\eta_s r}\,
   \cos\!\big[2(\alpha-\theta)\big]\,
   \hat{\mathbf r},
\qquad (\kappa r\ll1),
\label{eq:v_near}
\end{equation}
where $\alpha$ denotes the fixed orientation angle of the dipole axis
$\hat{\mathbf d}=(\cos\alpha,\sin\alpha)$ measured with respect to the
$x$-axis, while $\theta$ is the polar angle of the observation point
$\mathbf r=r(\cos\theta,\sin\theta)$. Thus, the velocity depends only on the
relative angle $\alpha-\theta$ between the dipole axis and the line of centers.
Equation~\eqref{eq:v_near} is characteristic of a two-dimensional stresslet
with logarithmic Stokeslet flow. An incompressible two-dimensional flow may be written in terms of a scalar stream function $\Psi(r,\theta)$ via
\begin{equation}
v_r = \frac{1}{r}\,\partial_\theta\Psi,
\qquad
v_\theta = -\partial_r\Psi.
\end{equation}
For Eq.~\eqref{eq:v_near} one finds
\begin{equation}
\Psi_{\rm near}(r,\theta)
 = -\frac{\sigma}{8\pi\eta_s}
   \sin\!\big[2(\alpha-\theta)\big],
\label{eq:psi_near}
\end{equation}
and the scalar vorticity,
\begin{equation}
\omega_z^{\rm (near)}(r,\theta)
 = -\frac{\sigma}{2\pi\eta_s r^2}
   \sin\!\big[2(\alpha-\theta)\big],
   \label{radvort}
\end{equation}
is nonzero except along the principal axes, indicating that the near-field flow is locally vortical. Fig.~\ref{fig:dipole_near_far} illustrates the flow.
\subsubsection{Exactly solvable quenched two-dipole dynamics in the near field}
We now show that the near-field velocity~\eqref{eq:v_near} leads to a simple and exactly solvable interaction for a pair of dipoles with \emph{quenched}
orientations.  Let us consider two dipoles with positions $\mathbf r_1,\mathbf r_2$,
strengths $\sigma_1,\sigma_2$, and fixed orientation angles
$\alpha_1,\alpha_2$, see Fig.~\ref{fig:dipole_schematic}.  Defining the relative coordinate
\[
\mathbf R=\mathbf r_1-\mathbf r_2 = r\,\hat{\mathbf r},
\qquad
\hat{\mathbf r}=(\cos\theta,\sin\theta),
\]
the induced velocities obtained from Eq.~\eqref{eq:v_near} yield the 
relative dynamics
\begin{equation}
\dot{\mathbf R}
=
\frac{1}{4\pi\eta_s r}
\Big[
\sigma_1\cos 2(\alpha_1-\theta)
+
\sigma_2\cos 2(\alpha_2-\theta)
\Big]\hat{\mathbf r}.
\label{eq:Rdot_near}
\end{equation}
The motion is therefore purely radial, implying
\begin{equation}
\dot\theta=0,
\end{equation}
so that the line of centers remains fixed.  Writing $\theta=\theta_0$,
the separation obeys
\begin{equation}
\frac{d}{dt}(r^2)
=
\frac{1}{2\pi\eta_s}
\Big[
\sigma_1\cos 2(\alpha_1-\theta_0)
+
\sigma_2\cos 2(\alpha_2-\theta_0)
\Big],
\end{equation}
which integrates to the exact solution
\begin{equation}
r^2(t)
=
r_0^2+
\frac{t}{2\pi\eta_s}
\Big[
\sigma_1\cos 2(\alpha_1-\theta_0)
+
\sigma_2\cos 2(\alpha_2-\theta_0)
\Big].
\label{eq:r2_exact}
\end{equation}
Thus, the quenched two-dipole problem in the near field is exactly solvable,
with the squared separation evolving linearly in time.  For identical
co-aligned dipoles, $\sigma_1=\sigma_2=\sigma$ and $\alpha_1=\alpha_2=\alpha$,
this reduces to
\begin{equation}
r^2(t)=r_0^2+\frac{\sigma}{\pi\eta_s}\cos 2(\alpha-\theta_0)\,t,
\end{equation}
implying separation for $\sigma>0$ and finite-time collapse for $\sigma<0$
depending on the relative alignment.
We list several special cases below,

(i) \emph{Collinear alignment} ($\theta_0=\alpha$):  
\begin{equation}
r^2(t)=r_0^2+\frac{\sigma}{\pi\eta_s}\,t.
\end{equation}
Pushers ($\sigma>0$) separate monotonically, while pullers ($\sigma<0$) undergo finite-time collapse at 
\begin{equation}
t_c=\frac{\pi\eta_s}{|\sigma|}\,r_0^2.
\end{equation}

(ii) \emph{Transverse configuration} ($\theta_0=\alpha+\pi/2$):  
\begin{equation}
r^2(t)=r_0^2-\frac{\sigma}{\pi\eta_s}\,t,
\end{equation}
so that the roles of pushers and pullers are reversed: pushers collapse and pullers separate.

(iii) \emph{Critical orientation} ($\theta_0=\alpha\pm\pi/4$):  
\begin{equation}
\cos 2(\alpha-\theta_0)=0,
\end{equation}
and the leading-order interaction vanishes, yielding
\begin{equation}
r(t)=r_0.
\end{equation}
These cases highlight the angular dependence of the dipolar interaction: the sign and strength of the radial drift are controlled entirely by the relative angle $\alpha-\theta_0$, leading to attraction, repulsion, or neutral behavior. Figure~\ref{fig:nearfield_verification} (Appendix~\ref{numver}) confirms these near-field predictions: the motion remains collinear, and $r^{2}(t)$ follows the exact linear evolution.
\subsection{Far field: screened hydrodynamics}
At distances much larger than the screening length, $\kappa r\gg1$, the exponential decay of $K_n(\kappa r)$ leaves only the algebraic terms in Eqs.~\eqref{eq:A_full}-\eqref{eq:B_full},
\begin{equation}
A(r)\simeq-\frac{1}{\kappa^2 r^2},
\qquad
B(r)\simeq \frac{2}{\kappa^2 r^2}.\nn
\end{equation}
Substituting into Eq.~\eqref{eq:dipolar_def} gives the far-field velocity
\begin{equation}
\mathbf v_{\rm far}(\mathbf r)
 = \frac{\sigma}{\pi\zeta_\parallel r^3}
   \Big[
     \cos\!\big(2\Delta\big)\,\hat{\mathbf r}
    -\sin\!\big(2\Delta\big)\,\hat{\boldsymbol\theta}
   \Big],
\qquad (\kappa r\gg1),
\label{eq:v_far}
\end{equation}
where we have defined the mismatch angle $\Delta \equiv \alpha-\theta$ and we have used $\kappa^2=\zeta_\parallel/\eta_s$.  The flow now decays as $r^{-3}$ and has both radial and azimuthal components.  The stream-function is
\begin{equation}
\Psi_{\rm far}(r,\theta)
 = -\frac{\sigma}{2\pi\zeta_\parallel}
    \frac{\sin\!\big[2(\alpha-\theta)\big]}{r^2},
\label{eq:psi_far}
\end{equation}
which  matches the near-field expression at $r\sim\kappa^{-1}$.  
In contrast to the near zone, the far-field flow is irrotational away from the dipole singularity:
\begin{equation}
\omega_z^{\rm (far)}(r,\theta)=0 \quad (r>0).
\label{farvort}
\end{equation}

\subsubsection{Exactly solvable two-dipole dynamics in the far field}
In the far field, the dipole dynamics is naturally quenched: since the flow is irrotational ($\omega_z=0$ for $r>0$), no hydrodynamic torque acts on the dipole axes, and their orientations remain fixed, in contrast to the vortical near-field regime. We now show that the two-dipole dynamics remains integrable in the far-field regime. We  consider the interaction of two force dipoles where the membrane flow is governed by the far-field velocity (re-written for convenience)
\begin{equation}
\mathbf v(\mathbf r)
=
\frac{\sigma}{\pi \zeta_\parallel r^3}
\Big[
\cos\!\big(2(\alpha-\theta)\big)\,\hat{\mathbf r}
-
\sin\!\big(2(\alpha-\theta)\big)\,\hat{\boldsymbol\theta}
\Big],
\label{eq:v_far_append}
\end{equation}
with the mismatch angle $\Delta=\alpha-\theta$. In contrast to the near field, the flow is
irrotational for $r>0$, so dipole orientations remain fixed and the dynamics reduces to a genuinely position-based problem (unlike the near-field where we had to enforce quenching). Let $\mathbf R=\mathbf r_1-\mathbf r_2$ denote the relative separation,
with polar coordinates $(r,\theta)$, see Fig.~\ref{fig:dipole_schematic}. For two dipoles with strengths
$\sigma_1,\sigma_2$ and fixed orientations $\alpha_1,\alpha_2$, the relative motion follows
\begin{align}
\dot r &=
\frac{1}{\pi \zeta_\parallel r^3}
\Big[
\sigma_1\cos\!\big(2(\alpha_1-\theta)\big)
+
\sigma_2\cos\!\big(2(\alpha_2-\theta)\big)
\Big],
\nn
\\
\dot\theta &=
-\frac{1}{\pi \zeta_\parallel r^4}
\Big[
\sigma_1\sin\!\big(2(\alpha_1-\theta)\big)
+
\sigma_2\sin\!\big(2(\alpha_2-\theta)\big)
\Big].
\label{eq:thetadot_general}
\end{align}
To simplify the dynamics, it is helpful to introduce the combinations
\begin{equation}
C = \sigma_1\cos 2\alpha_1 + \sigma_2\cos 2\alpha_2,
\qquad
S = \sigma_1\sin 2\alpha_1 + \sigma_2\sin 2\alpha_2,\nn
\end{equation}
in terms of which the dynamics can be recast as
\beqa
\dot r = \frac{1}{\pi \zeta_\parallel r^3}\big[C\cos 2\theta + S\sin 2\theta\big],\\
\dot\theta = -\frac{1}{\pi \zeta_\parallel r^4}\big[S\cos 2\theta - C\sin 2\theta\big].\nn
\eeqa
Defining $M=\sqrt{C^2+S^2}$ and $\phi=\tfrac12\arg(C+iS)$, and introducing the
angle $u=\theta-\phi$, the system reduces to
\begin{equation}
\dot r = \frac{M}{\pi \zeta_\parallel r^3}\cos(2u),
\qquad
\dot u = \frac{M}{\pi \zeta_\parallel r^4}\sin(2u).
\label{eq:reduced_far}
\end{equation}
Equations~\eqref{eq:reduced_far} are exactly integrable. The equations also reveal the invariant
\begin{equation}
\frac{\sin(2u)}{r^2} = \Lambda,
\label{eq:first_integral_far}
\end{equation}
with $\Lambda$ set by initial conditions. Using this relation, we find
\begin{equation}
\frac{d}{dt}\cos(2u)
=
-\frac{2M\Lambda^2}{\pi \zeta_\parallel},
\end{equation}
so that $\cos(2u)$ evolves linearly in time. The full solution is therefore
obtained by quadrature, with $r(t)$ tied to $u(t)$ through
Eq.~\eqref{eq:first_integral_far}. Unlike the near field, the dynamics are
intrinsically two-dimensional, with coupled radial and angular evolution.  
A simple limit is obtained for equal-strength, co-aligned dipoles,
$\sigma_1=\sigma_2=\sigma$ and $\alpha_1=\alpha_2=\alpha$. In this case
$M=2\sigma$ and $\phi=\alpha$, so that 
$u=\theta-\alpha=-\Delta$, with $\Delta=\alpha-\theta$. The reduced system
\eqref{eq:reduced_far} then simplifies to
\begin{equation}
\dot r=
\frac{2\sigma}{\pi \zeta_\parallel r^3}\cos(2\Delta),
\qquad
\dot\Delta=
\frac{2\sigma}{\pi \zeta_\parallel r^4}\sin(2\Delta).
\label{eq:coaligned_far}
\end{equation}
The invariant \eqref{eq:first_integral_far} becomes
\begin{equation}
\frac{\sin(2\Delta)}{r^2}=\Lambda,
\qquad
\Lambda=\frac{\sin(2\Delta_0)}{r_0^2},
\end{equation}
and the linear evolution law reduces to
\begin{equation}
\frac{d}{dt}\cos(2\Delta)
=
-\frac{4\sigma\Lambda^2}{\pi \zeta_\parallel}.
\end{equation}
Thus $\cos(2\Delta)$ varies linearly in time, while $r(t)$ follows from the
constraint $\sin(2\Delta)=\Lambda r^2$, yielding an exact parametric solution.
In contrast to the near-field case, the motion is generically non-collinear,
with a coupled evolution of separation and orientation. In the special
collinear configuration $\Delta_0=0$, the angular dynamics freezes and the
system reduces to purely radial motion,
\begin{equation}
r^4(t)=r_0^4+\frac{8\sigma}{\pi \zeta_\parallel}(t-t_0),
\end{equation}
that contrasts with the linear
$r^2(t)$ evolution in the near field.
We now consider special cases in the screened far-field regime.  

(i) \emph{Collinear alignment} ($\theta_0=\alpha$):  
Here $\Delta_0=0$, so $\sin(2\Delta_0)=0$ and the invariant
$\sin(2\Delta)/r^2=\Lambda$ gives $\Lambda=0$, implying $\Delta(t)=0$ for all time.  
The motion remains purely radial, with
\begin{equation}
\frac{d}{dt}r^4=\frac{8\sigma}{\pi\zeta_\parallel},
\qquad
r^4(t)=r_0^4+\frac{8\sigma}{\pi\zeta_\parallel}\,t.
\end{equation}
Pushers ($\sigma>0$) separate monotonically, while pullers ($\sigma<0$) undergo finite-time collapse at 
\begin{equation}
t_c=\frac{\pi\zeta_\parallel}{8|\sigma|}\,r_0^4.
\end{equation}

(ii) \emph{Transverse configuration} ($\theta_0=\alpha+\pi/2$):  
Here $\Delta_0=-\pi/2$, so again $\sin(2\Delta_0)=0$ and $\Lambda=0$, yielding $\Delta(t)=-\pi/2$.  
The radial dynamics becomes
\begin{equation}
\frac{d}{dt}r^4=-\frac{8\sigma}{\pi\zeta_\parallel},
\qquad
r^4(t)=r_0^4-\frac{8\sigma}{\pi\zeta_\parallel}\,t,
\end{equation}
so that the roles of pushers and pullers are reversed: pushers collapse in finite time, while pullers separate monotonically.

(iii) \emph{Critical orientation} ($\theta_0=\alpha\pm\pi/4$):  
\begin{equation}
\cos 2(\alpha-\theta_0)=0,
\end{equation}
so that the initial radial drift vanishes, $\dot r(0)=0$.  
Crucially, in contrast to the near field, the angular dynamics remains nonzero,
\begin{equation}
\dot\Delta(0)=\pm \frac{2\sigma}{\pi\zeta_\parallel r_0^4},
\end{equation}
so the system immediately rotates away from the critical orientation.  
Using the invariant $\sin(2\Delta)/r^2=\Lambda$ with $\Lambda=\pm r_0^{-2}$, together with the linear evolution law
\begin{equation}
\frac{d}{dt}\cos(2\Delta)
=
-\frac{4\sigma\Lambda^2}{\pi\zeta_\parallel},
\end{equation}
we obtain
\begin{equation}
\cos(2\Delta(t))
=
-\frac{4\sigma}{\pi\zeta_\parallel r_0^4}\,t,
\end{equation}
and hence, using $\sin^2(2\Delta)+\cos^2(2\Delta)=1$ together with $\sin(2\Delta)=\Lambda r^2$,
\begin{equation}
r^4(t)
=
r_0^4
-
\frac{16\sigma^2}{\pi^2\zeta_\parallel^2 r_0^4}\,t^2.
\end{equation}
Thus the critical configuration is not stationary: this branch reaches $r=0$ at the finite time
\begin{equation}
t_c=\frac{\pi\zeta_\parallel}{4|\sigma|}\,r_0^4.
\end{equation}

These cases again reflect the  angular structure of the dipolar interaction. However, in contrast to the near field, the vanishing of the radial drift at the critical orientation does not imply neutral dynamics: the nonzero azimuthal flow induces rotation of the line of centers, making the far-field dynamics intrinsically two-dimensional. As shown in Fig.~\ref{fig:farfield_verification} (Appendix~\ref{numver}), the far-field dynamics agrees with the analytic solution, including the linear evolution of $\cos(2u)$.

\section{Hamiltonian formulation for dipoles}
\label{sec:hamiltonian}

The \emph{incompressible} membrane enjoys a useful simplification: for force dipoles with \emph{fixed} orientations, the positional dynamics can be written in Hamiltonian form.  
Let the position of dipole $i$ be $\mathbf r_i=(x_i,y_i)$, with strength $\sigma_i=\pm1$ (pusher or puller) and fixed orientation angle $\alpha_i$.  
The velocity of dipole $i$ is obtained by evaluating the dipolar flow generated by all other dipoles $j\neq i$ at $\mathbf r_i$:
\begin{equation}
\dot{\mathbf r}_i
 = \sum_{j\neq i}
   \mathbf v_j(\mathbf r_i-\mathbf r_j).\nn
\end{equation}
Using the stream-function representation, the positional dynamics of the quenched dipoles can be written as
\begin{equation}
\dot{\mathbf r}_i
=
\sum_{j\neq i}
\hat{\mathbf z}\times \nabla_i \Psi_j(\mathbf r_{ij}),
\qquad
\mathbf r_{ij}=\mathbf r_i-\mathbf r_j,
\label{eq:stream_form_general}
\end{equation}
where $\Psi_j$ is the stream function generated by dipole $j$ (near- or far-field, depending on the regime of interest).  
Equation~\eqref{eq:stream_form_general} has the form of a planar Hamiltonian dynamics: the velocity of each dipole is obtained by rotating the gradient of an effective stream function generated by the remaining dipoles. Equivalently,
\begin{equation}
\sigma_i \dot{\mathbf r}_i
=
\hat{\mathbf z}\times
\left[
\sigma_i \sum_{j\neq i}\nabla_i \Psi_j(\mathbf r_{ij})
\right],
\label{eq:stream_form_general_sigma}
\end{equation}
which suggests a noncanonical Hamiltonian structure with $(x_i,y_i)$ as conjugate variables up to the weight $\sigma_i$. A simple scalar position-only Hamiltonian, however, emerges in closed form only under additional symmetry restrictions, which we now discuss.
\subsection{Near-field Hamiltonian}
As noted in Eq.~\eqref{radvort}, the unscreened near-field dipolar flow carries finite vorticity, so freely rotating dipoles undergo hydrodynamic
reorientation. These orientational dynamics preclude a Hamiltonian written
solely in terms of particle positions. A reduced Hamiltonian description is
nevertheless recovered in the \emph{quenched-orientation} limit, where dipole axes are externally constrained or relax on timescales slow compared to the translational dynamics. This regime is natural for anchored ``shakers’’ and for strongly aligned active suspensions~\cite{Shoham2023}. With orientations fixed, the near-field stresslet retains its $1/r$ form.
However, a consistent position-only Hamiltonian arises in a simple form only
in the \emph{co-aligned} quenched limit, $\alpha_i=\alpha$ for all $i$.
In this case, interactions are pairwise and the Hamiltonian is 
\begin{equation}
H_{\rm near}
=
\frac{1}{8\pi\eta_s}
\sum_{i<j}
\sigma_i\sigma_j
\sin\!\big[2(\theta_{ij}-\alpha)\big].
\label{eq:H_near}
\end{equation}
A global rotation sets $\alpha=0$, leaving a purely  angular
dependence. Differentiation of Eq.~\eqref{eq:H_near} yields the $1/r$ stresslet flows governing the near-field dynamics. In Cartesian coordinates, with $\Delta x_{ij}=x_i-x_j$ and
$\Delta y_{ij}=y_i-y_j$, Eq.~\eqref{eq:H_near} can be written as
\begin{equation}
H_{\rm near}
=
\frac{1}{8\pi\eta_s}
\sum_{i<j}
\sigma_i\sigma_j
\frac{
2\Delta x_{ij}\Delta y_{ij}\cos 2\alpha
-
(\Delta x_{ij}^2-\Delta y_{ij}^2)\sin 2\alpha
}{
\Delta x_{ij}^2+\Delta y_{ij}^2
}.
\label{eq:H_near_cart}
\end{equation}
\subsection{Far-field Hamiltonian}
In the screened regime ($\kappa r\gg1$), the dipolar flow becomes 
\emph{vorticity-free}, as shown in Eq.~\eqref{farvort}.  
In the absence of hydrodynamic torque, dipole orientations remain fixed, so the dipoles are \textit{naturally} quenched.  
A position-only Hamiltonian therefore exists.  
For co-aligned quenched dipoles, the screened stream function~\eqref{eq:psi_far} then yields
\begin{equation}
H_{\rm far}
=
\frac{1}{2\pi\zeta_\parallel}
\sum_{i<j}
\sigma_i\sigma_j
\frac{\sin\!\big[2(\theta_{ij}-\alpha)\big]}{r_{ij}^2}.
\label{eq:H_far}
\end{equation}
In Cartesian coordinates, with $\Delta x_{ij}=x_i-x_j$ and
$\Delta y_{ij}=y_i-y_j$, Eq.~\eqref{eq:H_far} becomes
\begin{equation}
H_{\rm far}
=
\frac{1}{2\pi\zeta_\parallel}
\sum_{i<j}
\sigma_i\sigma_j
\frac{
2\Delta x_{ij}\Delta y_{ij}\cos 2\alpha
-
(\Delta x_{ij}^2-\Delta y_{ij}^2)\sin 2\alpha
}{
(\Delta x_{ij}^2+\Delta y_{ij}^2)^2
}.
\label{eq:H_far_cart}
\end{equation}
The associated equations of motion,
\begin{equation}
\dot{x}_i = \frac{1}{\sigma_i} \frac{\partial H}{\partial y_i},
\qquad
\dot{y}_i = -\frac{1}{\sigma_i} \frac{\partial H}{\partial x_i},
\label{eq:ham_eom}
\end{equation}
generate the screened $1/r^3$ dipolar flows in the co-aligned quenched
limit. For co-aligned dipoles, a global rotation sets $\alpha=0$, and both the near- and far-field Hamiltonians reduce to simple  forms,
\begin{equation}
H \;\propto\;
\sum_{i<j}
\sigma_i\sigma_j
\frac{\sin(2\theta_{ij})}{r_{ij}^n},
\qquad
n=0 \ \text{(near)},\qquad n=2 \ \text{(far)}.
\end{equation}
\section{Two-body Hamiltonian in relative variables}
\label{ppanalysis}
\begin{figure}[t]
\centering
\includegraphics[width=\linewidth]{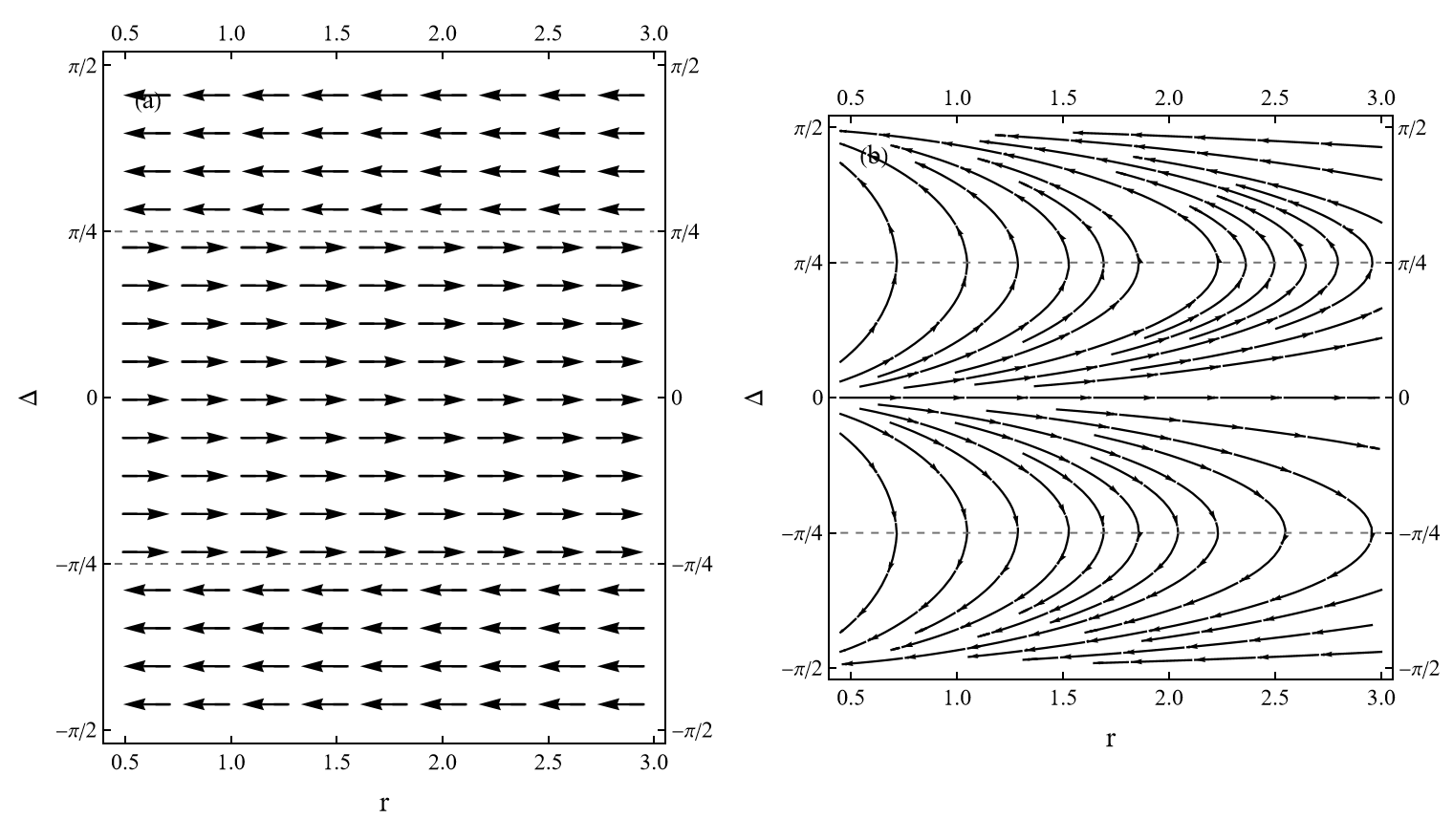}
\caption{
Phase portraits of the reduced co-aligned two-dipole dynamics in a supported incompressible membrane, shown in $(r,\Delta)$ with $\Delta=\alpha-\theta$ for equal-strength pushers. 
(a) Near field: purely radial flow with $\dot\Delta=0$, so the nullclines $\Delta=\pm\pi/4$ are stationary and separate attractive and repulsive sectors. 
(b) Far field: screened dynamics with coupled radial and angular motion; the same nullclines remain radial turning lines but are no longer stationary, reflecting the intrinsically two-dimensional phase-space flow.
}
\label{fig:phaseportrait_pushers_supported}
\end{figure}
To further elucidate the Hamiltonian structure, it is useful to examine the reduced two-body problem in relative coordinates. In this case, the dynamics simplifies substantially and admits an explicit Hamiltonian formulation in terms of a single pair of canonical variables. This reduction provides a framework in which the qualitative differences between the near- and far-field interactions become manifest, and allows for a direct comparison with the phase-space structure shown in Fig.~\ref{fig:phaseportrait_pushers_supported}. For the finite-subphase supported membrane, the quenched near-zone two-dipole dynamics also admits a simple Hamiltonian reduction. Writing the relative coordinate as
\[
\mathbf R=\mathbf r_1-\mathbf r_2=r\,\hat{\mathbf r},
\qquad
\hat{\mathbf r}=(\cos\theta,\sin\theta),
\]
the near-field flow gives
\[
\dot r
=
\frac{1}{4\pi\eta_s r}
\Big[
\sigma_1\cos 2(\alpha_1-\theta)
+
\sigma_2\cos 2(\alpha_2-\theta)
\Big],
\qquad
\dot\theta=0.
\]
Introducing canonical variables
\[
q=\theta,
\qquad
p=-\frac{r^2}{2},
\]
we find
\[
\dot q=\frac{\partial H}{\partial p}=0,
\qquad
\dot p=-\frac{\partial H}{\partial q},
\]
generated by
\begin{equation}
H_{\rm near}^{(2)}(q)
=
-\frac{1}{8\pi\eta_s}
\Big[
\sigma_1\sin 2(\alpha_1-q)
+
\sigma_2\sin 2(\alpha_2-q)
\Big].
\end{equation}
Because the Hamiltonian is independent of $p$, the angular variable is frozen and the motion is effectively one-dimensional. The radial evolution then integrates exactly to
\[
r^2(t)=r_0^2+
\frac{t}{2\pi\eta_s}
\Big[
\sigma_1\cos 2(\alpha_1-\theta_0)
+
\sigma_2\cos 2(\alpha_2-\theta_0)
\Big].
\]
For equal-strength, co-aligned dipoles, $\sigma_1=\sigma_2=\sigma$ and $\alpha_1=\alpha_2=\alpha$, this reduces to
\[
H_{\rm near}^{(2)}(q)= -\frac{\sigma}{4\pi\eta_s}\sin 2(\alpha-q),
\]
with $r^2(t)=r_0^2+\frac{\sigma}{\pi\eta_s}\cos 2(\alpha-\theta_0)t$. In the screened far field, the quenched two-dipole problem also admits a Hamiltonian formulation. Using the same canonical variables
\[
\mathbf R=\mathbf r_1-\mathbf r_2=r\,\hat{\mathbf r},
\qquad
q=\theta,
\qquad
p=-\frac{r^2}{2},
\]
the far-zone equations of motion become
\beqa
\dot r=
\frac{1}{\pi \zeta_\parallel r^3}
\Big[
\sigma_1\cos 2(\alpha_1-\theta)+
\sigma_2\cos 2(\alpha_2-\theta)
\Big],\nn\\
\dot\theta=
-\frac{1}{\pi \zeta_\parallel r^4}
\Big[
\sigma_1\sin 2(\alpha_1-\theta)+
\sigma_2\sin 2(\alpha_2-\theta)
\Big].
\eeqa
These are generated by the Hamiltonian
\begin{equation}
H_{\rm far}^{(2)}(q,p)
=
-\frac{1}{2\pi\zeta_\parallel r^2}
\Big[
\sigma_1\sin 2(\alpha_1-q)+
\sigma_2\sin 2(\alpha_2-q)
\Big],
\end{equation}
with $r=\sqrt{-2p}$ through $\dot q=\partial H/\partial p$ and $\dot p=-\partial H/\partial q$. Unlike the near-field Hamiltonian, which is independent of $p$, the screened far-field Hamiltonian depends explicitly on the radial coordinate through $r^{-2}$. This produces genuine radial-angular coupling and makes the two-body dynamics intrinsically two-dimensional, while preserving exact integrability. To visualize the reduced quenched two-dipole dynamics in the supported membrane, it is useful to examine the phase portraits in the plane of relative separation $r=|\mathbf r_1-\mathbf r_2|$ and relative orientation mismatch $\Delta=\alpha-\theta$, where $\alpha$ is the common fixed dipole axis and $\theta$ is the polar angle of the line of centers. The geometric definitions of the relative separation $r$ and angle $\theta$, as well as the dipole orientations $\alpha_i$, are illustrated in the schematic of Fig.~\ref{fig:dipole_schematic}. For equal-strength co-aligned dipoles, the dynamics reduces to a closed two-dimensional system for $(r,\Delta)$, with distinct forms in the near- and far-field limits. In the unscreened near field, the leading-order equations are
\[
\dot r=\frac{\sigma}{2\pi\eta_s r}\cos(2\Delta),\qquad \dot\Delta=0,
\]
so the angular variable is frozen and the flow in phase space is purely horizontal. Figure~\ref{fig:phaseportrait_pushers_supported}(a) was constructed accordingly using a uniform horizontal arrow field whose direction is set by the sign of $\cos(2\Delta)$. The portrait immediately explains the near-field dipolar interaction: for pushers ($\sigma>0$), the region $|\Delta|<\pi/4$ corresponds to outward radial drift and hence monotonic separation, whereas the regions $\pi/4<|\Delta|<\pi/2$ correspond to inward drift and hence approach. The lines $\Delta=\pm\pi/4$ are radial nullclines where $\dot r=0$; because $\dot\Delta=0$ everywhere in the near-field problem, these are in fact entire stationary lines in the reduced phase space. The near-field portrait therefore makes explicit that the quenched dynamics are effectively one-dimensional despite being plotted in the two variables $(r,\Delta)$. In the screened far field, by contrast, the reduced equations become
\[
\dot r=\frac{2\sigma}{\pi\zeta_\parallel r^3}\cos(2\Delta),\qquad
\dot\Delta=\frac{2\sigma}{\pi\zeta_\parallel r^4}\sin(2\Delta),
\]
so radial and angular evolution are coupled. Figure~\ref{fig:phaseportrait_pushers_supported}(b) was obtained from the corresponding vector field using streamlines in the $(r,\Delta)$ plane. The same geometric factor $\cos(2\Delta)$ still controls the sign of the radial drift, so $\Delta=\pm\pi/4$ remain radial nullclines, shown as dashed lines. However, unlike in the near field, these are no longer stationary because $\dot\Delta\neq0$ there except at special points. Thus an initial condition placed on a critical line has vanishing instantaneous radial velocity but is nevertheless advected vertically in phase space, immediately rotating away from criticality. This is the geometric manifestation of the fact that the far-field dynamics are intrinsically two-dimensional. The line $\Delta=0$ is an invariant manifold corresponding to collinear alignment, and along it pushers separate monotonically. In contrast, trajectories beginning near $\Delta=\pm\pi/2$ lie in the transverse sector where $\cos(2\Delta)<0$, so the radial motion is inward as shown in Fig.~\ref{fig:phaseportrait_pushers_supported}(b). \\\\
The origin of aggregation in the screened regime can be understood from the reduced two-body dynamics in $(r,\Delta)$, where $r$ is the separation and $\Delta=\alpha-\theta$ the orientation mismatch angle (see Appendix~\ref{ppdetails} for more details). In the near field, $\dot\Delta=0$, so the dynamics is purely radial and phase space is partitioned by the nullclines $\Delta=\pm\pi/4$ into repulsive and attractive sectors that cannot be crossed; configurations that are initially non-attracting therefore remain non-aggregating. In contrast, in the far field the dynamics is intrinsically two-dimensional, with $\dot\Delta\neq0$ coupling angular and radial motion. This induces a systematic drift in $\Delta$ that allows trajectories to cross from the initially repulsive sector into the attractive one, providing a robust mechanism for aggregation. The flow admits an invariant $\sin(2\Delta)/r^2$, which constrains trajectories and selects a universal transverse collapse geometry with $\Delta\to\pm\pi/2$ as $r\to0$. Although the phase portrait suggests convergence toward small separation, the reduced far-field dynamics does not behave as a usual attractor. The system remains Hamiltonian for $r>0$, with the invariant $\sin(2\Delta)/r^2=\Lambda$, and hence evolves along phase-space curves without dissipation. The observed aggregation corresponds instead to finite-time collapse onto the singular boundary $r=0$, where the effective interaction potential diverges as $r^{-2}$. Near collapse we find $\dot r\sim -2\sigma/(\pi\zeta_\parallel r^3)$, giving $r(t)\sim (t_c-t)^{1/4}$ with $t_c$ being the collapse time. The invariant $\sin(2\Delta)/r^2=\Lambda$ further implies that the collapse geometry is selected by the angular dynamics: for pushers ($\sigma>0$), trajectories approach the transverse configurations $\Delta\to\pm\pi/2$, while for pullers ($\sigma<0$) the phase-space flow reverses and collapse occurs along the collinear manifold $\Delta\to 0$. These asymptotic behaviours are consistent with the phase portraits in Fig.~\ref{fig:phaseportrait_pushers_supported} and are confirmed numerically (see Appendix~\ref{ppdetails} and Appendix~\ref{supfig}).

\section{Collective dynamics of pusher and puller dipoles}
\label{sec:collective}
\begin{figure*}[t]
\centering

\begin{subfigure}[t]{0.23\linewidth}
    \centering
    \includegraphics[width=\linewidth]{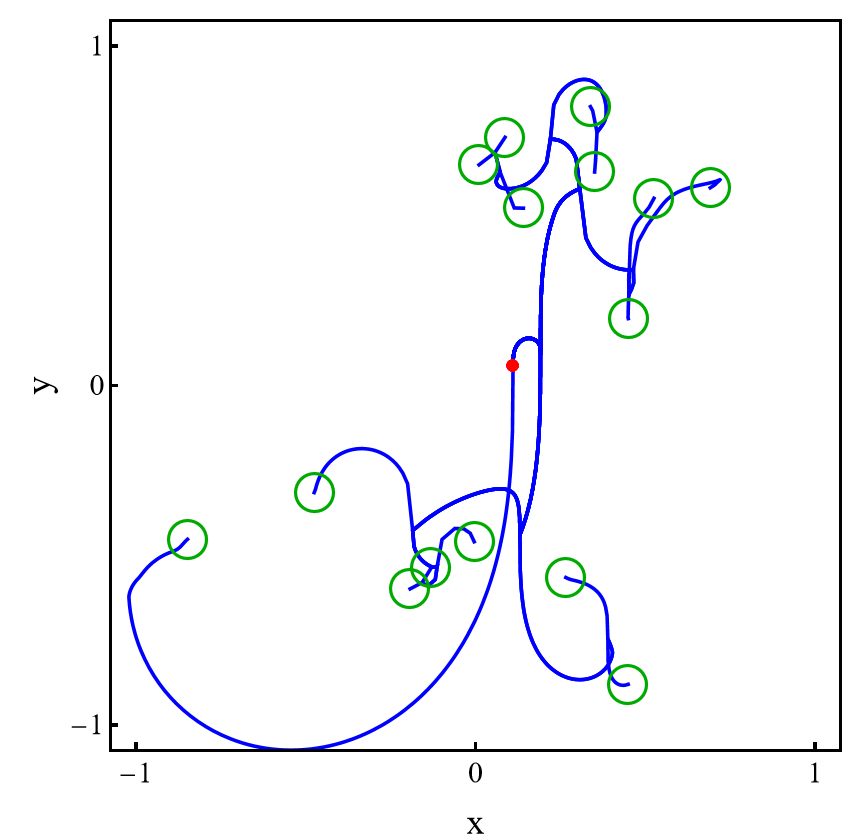}
    \caption{}
\end{subfigure}
\hfill
\begin{subfigure}[t]{0.23\linewidth}
    \centering
    \includegraphics[width=\linewidth]{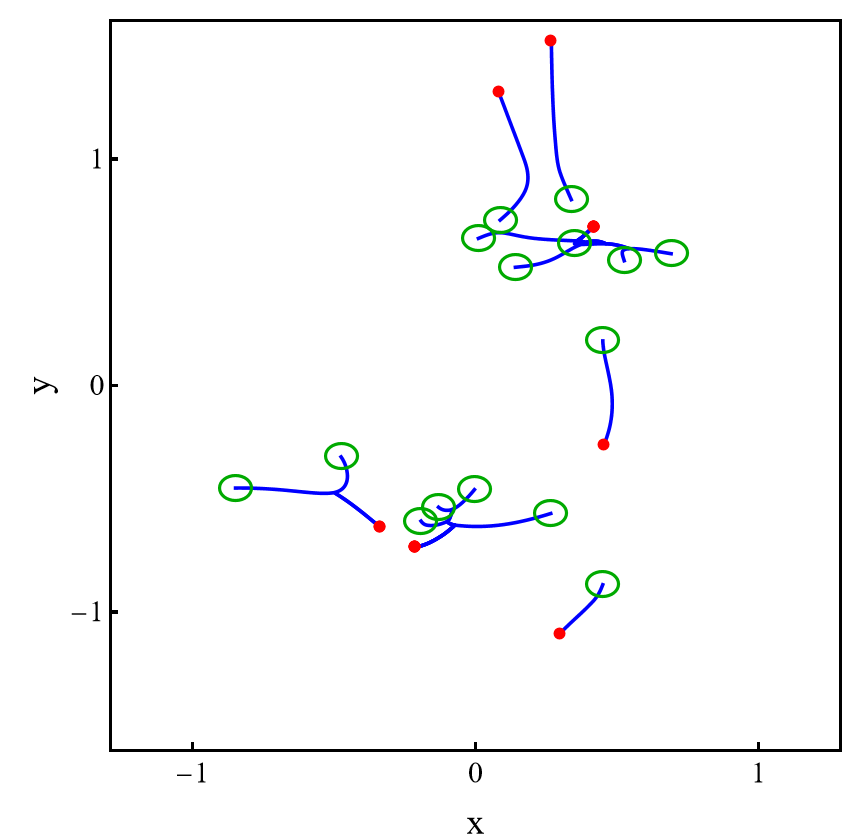}
    \caption{}
\end{subfigure}
\hfill
\begin{subfigure}[t]{0.23\linewidth}
    \centering
    \includegraphics[width=\linewidth]{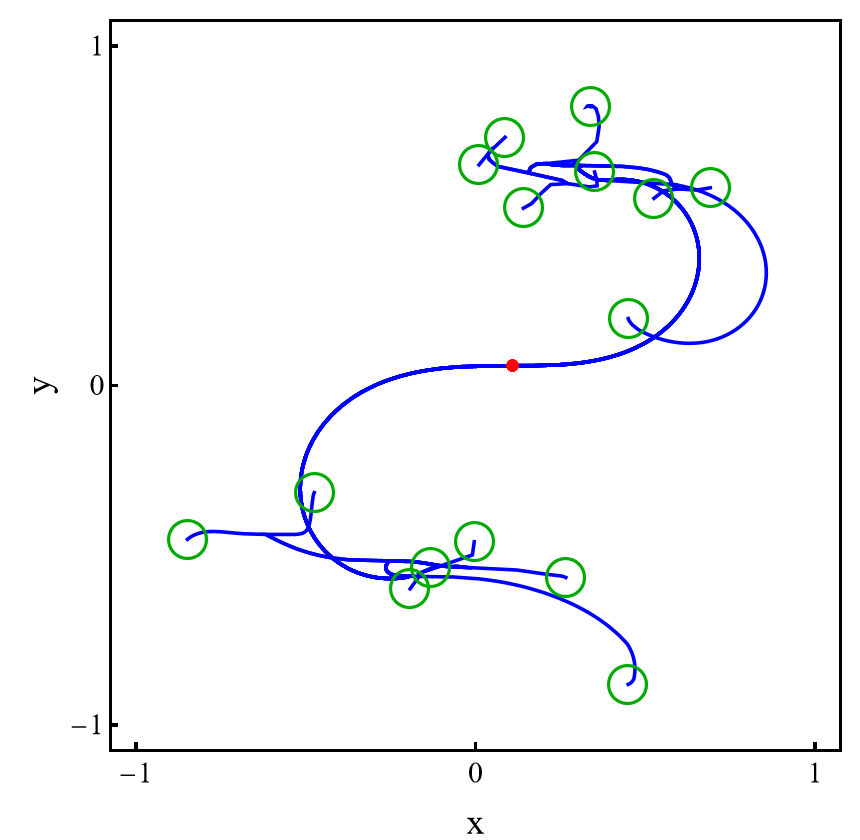}
    \caption{}
\end{subfigure}
\hfill
\begin{subfigure}[t]{0.23\linewidth}
    \centering
    \includegraphics[width=\linewidth]{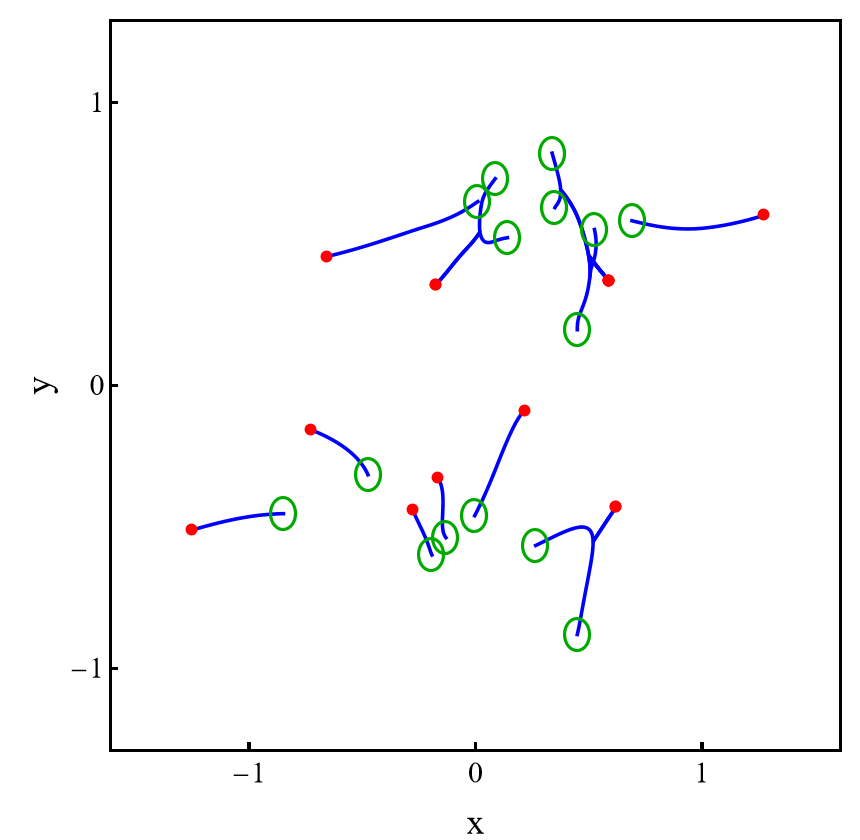}
    \caption{}
\end{subfigure}

\vspace{0.5em}

\begin{subfigure}[t]{0.23\linewidth}
    \centering
    \includegraphics[width=\linewidth]{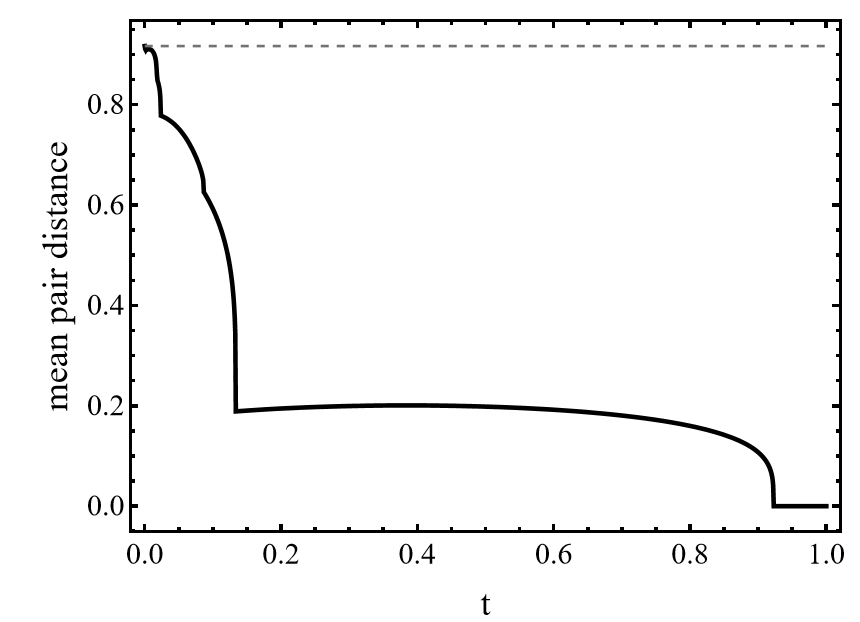}
    \caption{}
\end{subfigure}
\hfill
\begin{subfigure}[t]{0.23\linewidth}
    \centering
    \includegraphics[width=\linewidth]{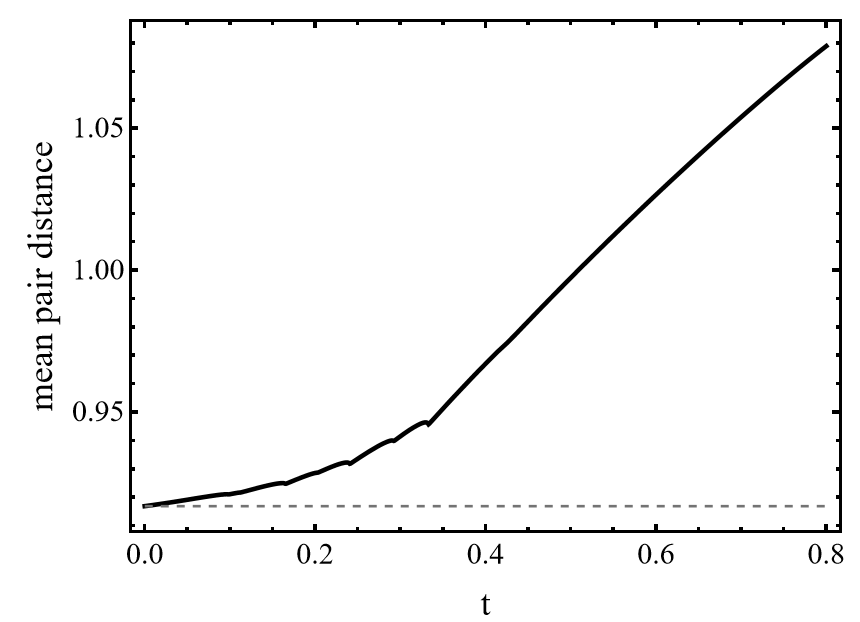}
    \caption{}
\end{subfigure}
\hfill
\begin{subfigure}[t]{0.23\linewidth}
    \centering
    \includegraphics[width=\linewidth]{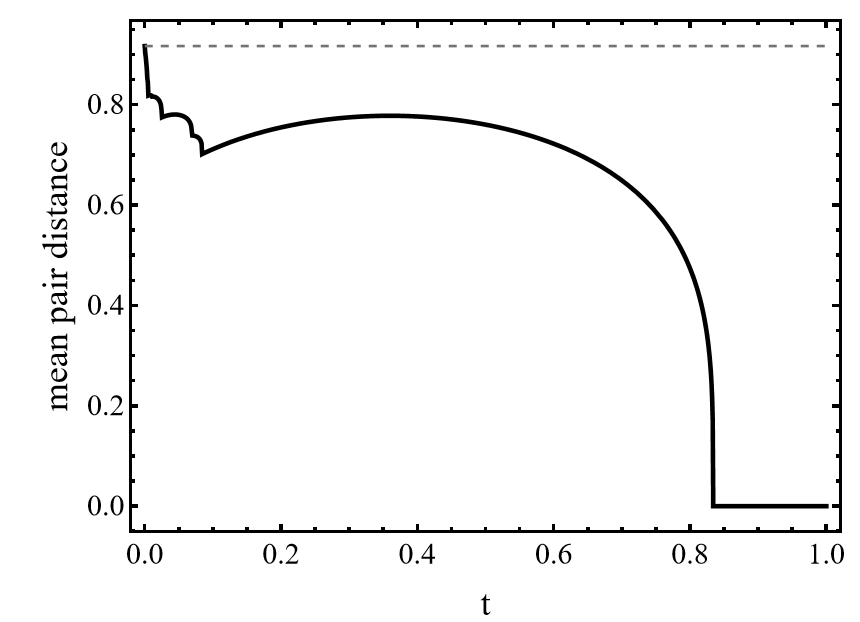}
    \caption{}
\end{subfigure}
\hfill
\begin{subfigure}[t]{0.23\linewidth}
    \centering
    \includegraphics[width=\linewidth]{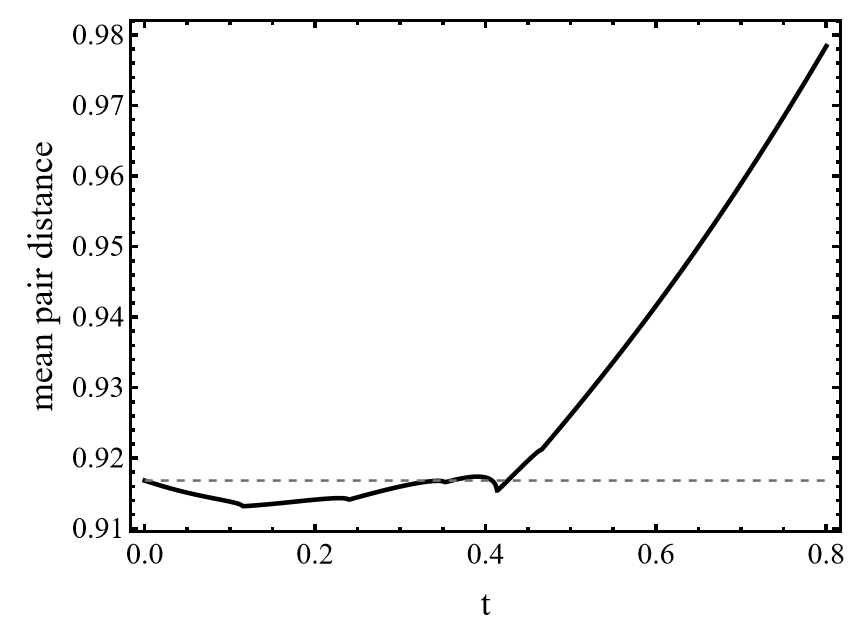}
    \caption{}
\end{subfigure}

\caption{
Collective dynamics of 15 co-aligned dipoles in an incompressible supported membrane. 
Top row: particle trajectories in the plane. Bottom row: corresponding mean pair distance as a function of time. 
From left to right: pusher (far field), pusher (near field), puller (far field), and puller (near field). 
In the trajectory panels, open green circles denote initial positions and filled red markers denote final positions. 
In the distance plots, the dashed horizontal line marks the initial mean pair distance. 
}
  \label{fig:combined_push_pull}
\end{figure*}
We now probe the dynamical consequences of the two Hamiltonians $H_{\rm far}$ and $H_{\rm near}$ by separately
time-evolving ensembles of identical pushers and pullers ($\sigma_i=\pm1$)
with fixed, co–aligned orientations ($\alpha=0$).  
In the far field, the flow is vorticity–free and orientations are
naturally quenched, making $H_{\rm far}$ an exact position–based
Hamiltonian.  
In the near field, where vorticity is finite, a Hamiltonian description
applies only in the enforced quenched–orientation limit.  
Within these regimes, we contrast the collective organization produced by
$H_{\rm far}$ and $H_{\rm near}$.
\subsection{Simulation details}
We integrate the $2N$ position variables $\{x_i(t),y_i(t)\}$ using a standard adaptive Runge-Kutta method, evolving Eq.~\eqref{eq:ham_eom} with either $H_{\rm near}$ or $H_{\rm far}$.  
Throughout this section we take $N=15$ and use identical random initial positions for both regimes: particles are sampled uniformly from a unit disc.  
Initial positions are indicated by open green circles in Fig.~\ref{fig:combined_push_pull}, while filled red circles mark the final positions at $t=t_{\rm max}$. To quantify aggregation or dispersion, we monitor the mean pairwise separation
\begin{equation}
\langle d_{ij}\rangle(t)
 = \frac{2}{N(N-1)}
   \sum_{i<j}
   \sqrt{\Delta x_{ij}^2+\Delta y_{ij}^2},
\end{equation}
where $\Delta x_{ij}=x_i-x_j$ and $\Delta y_{ij}=y_i-y_j$. A small softening parameter $\epsilon$ was added inside the squared distance to regularize close encounters without altering the large–scale flow. Energy conservation, $\dot H=0$ (between collisions), is verified to within a relative tolerance of $10^{-6}$, confirming numerical accuracy, similar to the investigations in Ref.~\cite{Shoham2023}.
A detailed account of the numerical implementation for both near- and far-field quenched dynamics, including the softening procedure and details of Hamiltonian conservation, is provided is provided in Appendices~\ref{app:numerics_near} and~\ref{app:numerics_far}. The short-distance regularization used here is implemented via a soft regulator in the velocity kernels (Appendices~\ref{app:numerics_near} and~\ref{app:numerics_far}); we have verified that replacing it with an explicit short-range harmonic repulsion leads to identical conclusions (Appendix~\ref{supfig}).
\subsection{Near vs.\ far: aggregation and dispersion}
The collective dynamics of active dipoles in an incompressible membrane depend
crucially on whether interactions are mediated by the screened Hamiltonian
$H_{\rm far}$ or the unscreened $H_{\rm near}$. Figure~\ref{fig:combined_push_pull}
summarizes the evolution of ensembles of $N=15$ co-aligned dipoles for both
pushers and pullers. 

Under the screened far-field Hamiltonian $H_{\rm far}$
[Figs.~\ref{fig:combined_push_pull}(a,c,e,g)], both motor types undergo rapid
collapse into compact aggregates: trajectories contract toward small
separations and the mean pair distance $\langle d_{ij}\rangle(t)$ decreases
sharply in time. In contrast, evolution under the unscreened near-field
Hamiltonian $H_{\rm near}$ [Figs.~\ref{fig:combined_push_pull}(b,d,f,h)]
exhibits no such collapse. Instead, the ensemble spreads into extended
configurations, with $\langle d_{ij}\rangle(t)$ increasing monotonically.

These contrasting behaviours are directly consistent with the reduced
two-body phase-space structure discussed in
Sec.~\ref{ppanalysis} and shown in
Fig.~\ref{fig:phaseportrait_pushers_supported}
(see also Appendix~\ref{ppdetails}).
In the near field, the dynamics is effectively one-dimensional with
$\dot\Delta=0$, so phase space is partitioned by the nullclines
$\Delta=\pm\pi/4$ into attractive and repulsive sectors that cannot be crossed.
As a result, initially non-attracting configurations remain non-aggregating at
the many-body level. 

In the far field, by contrast, the dynamics is intrinsically two-dimensional,
with $\dot\Delta\neq0$ coupling angular and radial motion. This induces a
systematic drift in $\Delta$ that allows trajectories to cross from
repulsive to attractive sectors, providing a robust mechanism for
aggregation. The resulting many-body collapse arises from angular transport across
phase space enabled by the screened hydrodynamic interaction.

The similarity between pusher and puller ensembles further highlights that
the dominant control parameter is the hydrodynamic kernel itself.
Screened interactions ($H_{\rm far}$) generically promote aggregation,
while unscreened interactions ($H_{\rm near}$) suppress collective collapse,
independently of dipole sign. A representative far-field evolution with arbitrary orientations
(Fig.~\ref{repr}) in the Appendix~\ref{supfig}  confirms that this mechanism persists beyond the
co-aligned limit. Additional plots of multi-particle collapse are provided in
Appendix~\ref{supfig}. This mechanism also distinguishes the nature of the collapse for pushers and pullers. 
For pushers ($\sigma>0$), the phase portrait in 
Fig.~\ref{fig:phaseportrait_pushers_supported} shows that trajectories are driven toward the transverse configurations $\Delta=\pm\pi/2$ as $r\to0$, yielding collapse with $\Delta\to\pm\pi/2$. 
For pullers ($\sigma<0$), the phase-space flow reverses sign, so that trajectories instead approach the collinear configuration $\Delta=0$ as $r\to0$. 
These distinct collapse configurations are clearly visible in the many-body simulations and in the representative trajectories shown in Appendix~\ref{supfig}. 
\section{Summary and outlook}
\label{sec:discussion}

We have developed an analytic framework for force-dipole hydrodynamics in an \emph{incompressible} supported membrane, based on the Brinkman-regularized membrane Stokes equation.  
Starting from the screened Green's tensor, we derived the corresponding dipolar flow field and identified its two asymptotic regimes: an unscreened near field with logarithmic Stokeslet structure and a screened far field with algebraically decaying interactions. The associated stream functions and vorticity fields show that these two regimes differ not only in spatial decay, but also in the angular structure: the near field is vortical, whereas the far field is irrotational away from the dipole singularity.

This distinction has important dynamical consequences.  
In the near field, the leading-order dipolar flow is purely radial. For two dipoles with quenched orientations, the relative angle is therefore conserved and the dynamics become exactly solvable, with the squared separation evolving linearly in time. In the far field, by contrast, the screened interaction contains both radial and azimuthal components. The resulting two-body problem remains exactly integrable, but is intrinsically two-dimensional: the dipole-dipole separation and line-of-centers angle evolve in a coupled manner, with their evolution linked via a conserved quantity.  In particular, the familiar collinear configuration reduces to a linear law for \(r^4(t)\).

A second main result is the emergence of Hamiltonian structure in the co-aligned quenched situations.  
For incompressible supported membranes, the positional dynamics can be written in Hamiltonian form, but the structure of the Hamiltonian is remarkably different in the two regimes. In the far field, because the flow is vorticity free, the dipoles are naturally quenched and yield an exact position-based Hamiltonian \(H_{\rm far}\) for co-aligned dipoles. In the near field, finite vorticity would in general induce orientational dynamics, so a position-only Hamiltonian is available only in the enforced quenched-orientation limit, giving an effective Hamiltonian \(H_{\rm near}\). In contrast, in the far field the Hamiltonian description is an exact one, due to absence of vorticity of the dipolar flows. At the two-body level, this difference is especially transparent: the near-field Hamiltonian is independent of the radial momentum variable and therefore produces effectively one-dimensional motion, whereas the far-field Hamiltonian depends explicitly on it and generates genuine radial-angular coupling.

These structural differences are reflected clearly in the collective dynamics.  
Numerical evolution of many co-aligned dipoles under the two Hamiltonians shows qualitatively distinct behavior. Under the screened far-field Hamiltonian \(H_{\rm far}\), both pushers and pullers exhibit a strong tendency toward clustering, with trajectories collapsing into compact aggregates and the mean pair separation decreasing sharply in time. Under the unscreened near-field Hamiltonian \(H_{\rm near}\), the same ensembles instead spread into extended configurations, and the mean pair separation grows monotonically. Thus, within the present incompressible supported-membrane setup, the key factor controlling collective organization is the nature of the screened hydrodynamic interaction rather than the sign of the dipoles.
The resulting many-body collapse in the time evolution of the far-field Hamiltonian  arises from angular transport across
phase space enabled by the screened hydrodynamic interaction.
Taken together, these results identify incompressible supported membranes as a particularly simple example in which explicit hydrodynamic Hamiltonians can be constructed and analyzed for active particles. They also show that hydrodynamic screening changes the geometry of the flow, the structure of the dynamical phase space, and the nature of many-body organization.

Several extensions remain of interest.  
Allowing membrane compressibility would introduce a longitudinal mobility channel and, in general, spoil the present position-only Hamiltonian structure. Likewise, odd-viscous or other non-dissipative active stresses could modify the phase-space geometry and potentially lead to generalized Hamiltonian or noncanonical formulations. It would also be natural to examine how curvature, confinement, thermal fluctuations, or biochemical activity alter the crossover between near- and far-field behavior established here. More broadly, the present results provide a useful starting point for connecting hydrodynamic models of membrane-bound active matter with experimental setups, where both the screening length and membrane viscosity can be tuned systematically.

\begin{acknowledgments}
It is a pleasure to thank Sarthak Bagaria, Naomi Oppenheimer, Haim Diamant, Michael D.\ Graham, and Mark Henle.   S.K.\ acknowledges support from an Institute fellowship at Birla Institute of Technology and Science, Pilani (Hyderabad Campus).  
R.S.\ is supported by a DST INSPIRE Faculty fellowship, India (Grant No.\ IFA19-PH231), NFSG, and an OPERA Research Grant from Birla Institute of Technology and Science, Pilani (Hyderabad Campus). 
\end{acknowledgments}
\section*{Data Availability}
The code and numerical data supporting the findings of this study are publicly available on Zenodo at Ref.~\cite{KrishnanSamantaCode}.

\appendix
\section{Numerical verification of analytic solutions for the dipole pair in the near- and far-field regimes}
\label{numver}
\begin{figure}[t]
\centering
\includegraphics[width=0.9\linewidth]{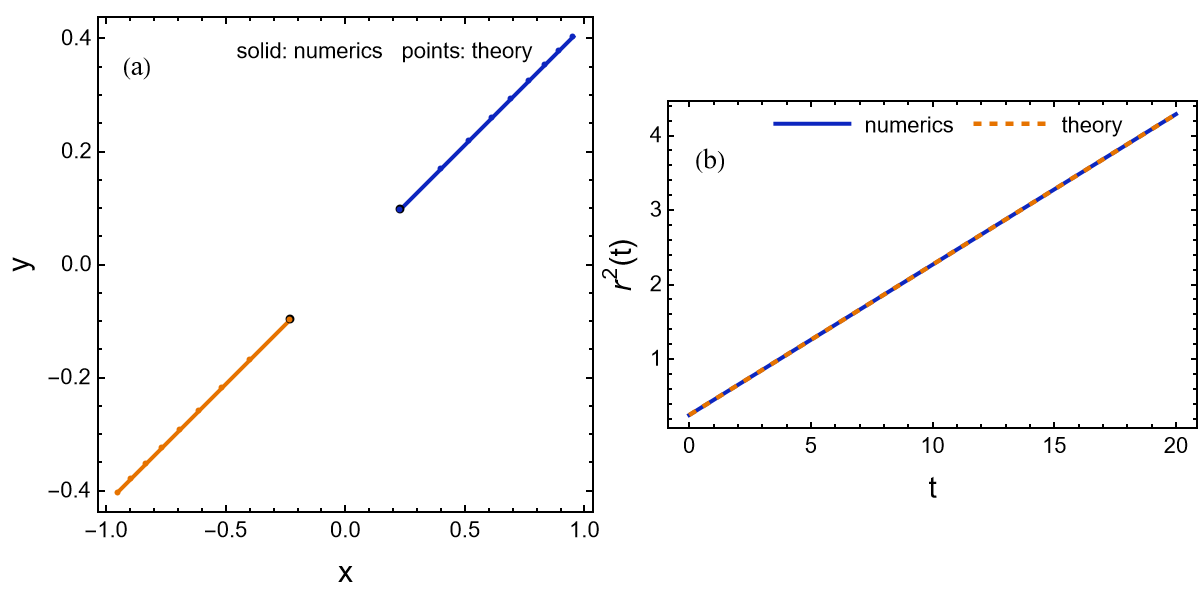}
\caption{Near-field dynamics verification for two quenched dipoles, for separating trajectories. (a) Numerical trajectories of the two dipoles in the plane (solid), with the exact analytic result overlaid as points.  (b) Comparison of the numerical solution for $r^2(t)$ (solid) with the exact prediction
$
r^2(t)=r_0^2+\frac{t}{2\pi\eta_s}
\Big[\sigma_1\cos 2(\alpha_1-\theta_0)+\sigma_2\cos 2(\alpha_2-\theta_0)\Big],
$ showing agreement. The parameters are $\eta_s=1$, $\sigma_1=\sigma_2=1$, $\alpha_1=\pi/7$, $\alpha_2=\pi/3$, $r_0=0.5$, and $\theta_0=0.4$.
}
\label{fig:nearfield_verification}
\end{figure}

\begin{figure}[t]
\centering
\includegraphics[width=0.48\linewidth]{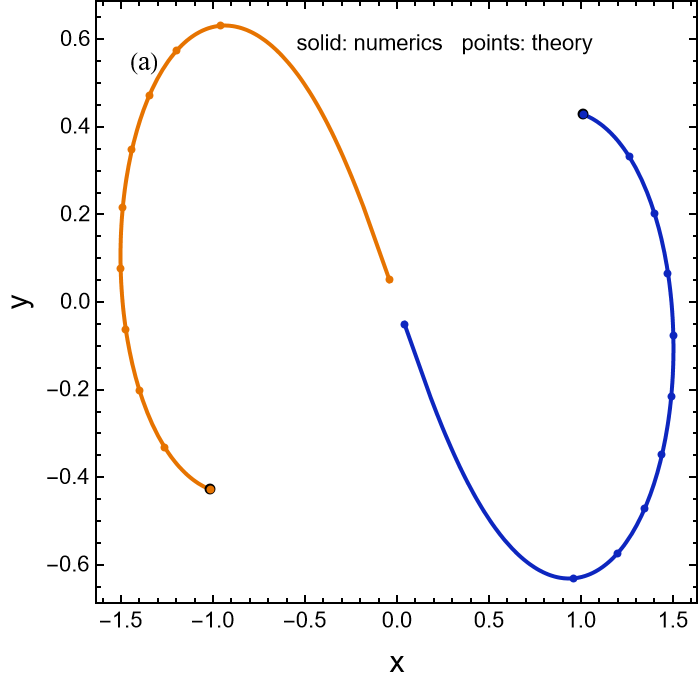}
\hfill
\includegraphics[width=0.48\linewidth]{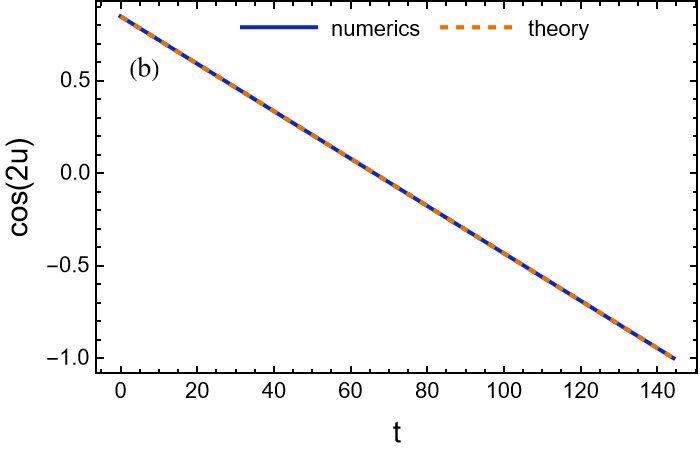}
\caption{Far-field dynamics of two  dipoles in a supported membrane ($\kappa r \gg 1$), showing agreement between numerical integration and the exact analytic solution. (a) Trajectories of the two dipoles in the plane: solid curves denote numerical solutions, while symbols indicate the analytic solution. (b) Verification of the exact linear evolution $\cos(2u(t))=\cos(2u_0)+\dot{\cos(2u)}\,t$, showing agreement between numerics (solid) and theory (dashed). Initial conditions and parameters are  $\sigma_1=1.2$, $\sigma_2=0.8$, 
$\alpha_1=\pi/7$, $\alpha_2=\pi/3$, 
$r_0=2.2$, $\theta_0=0.4$, with $\zeta_\parallel=1$. }
\label{fig:farfield_verification}
\end{figure}

In this appendix, we present numerical results that verify the analytic solutions derived in Sec.~\ref{sec:dipolar} of the main text. Specifically, we provide two representative plots demonstrating agreement between theory and numerics in both the near- and far-field regimes. Figure~\ref{fig:nearfield_verification} shows the near-field verification, while Fig.~\ref{fig:farfield_verification} presents the corresponding results in the far-field regime.

\section{Detailed analysis of near and far-field two-dipole phase portraits}
\label{ppdetails}
\begin{figure*}[t]
\centering
\includegraphics[width=0.95\linewidth]{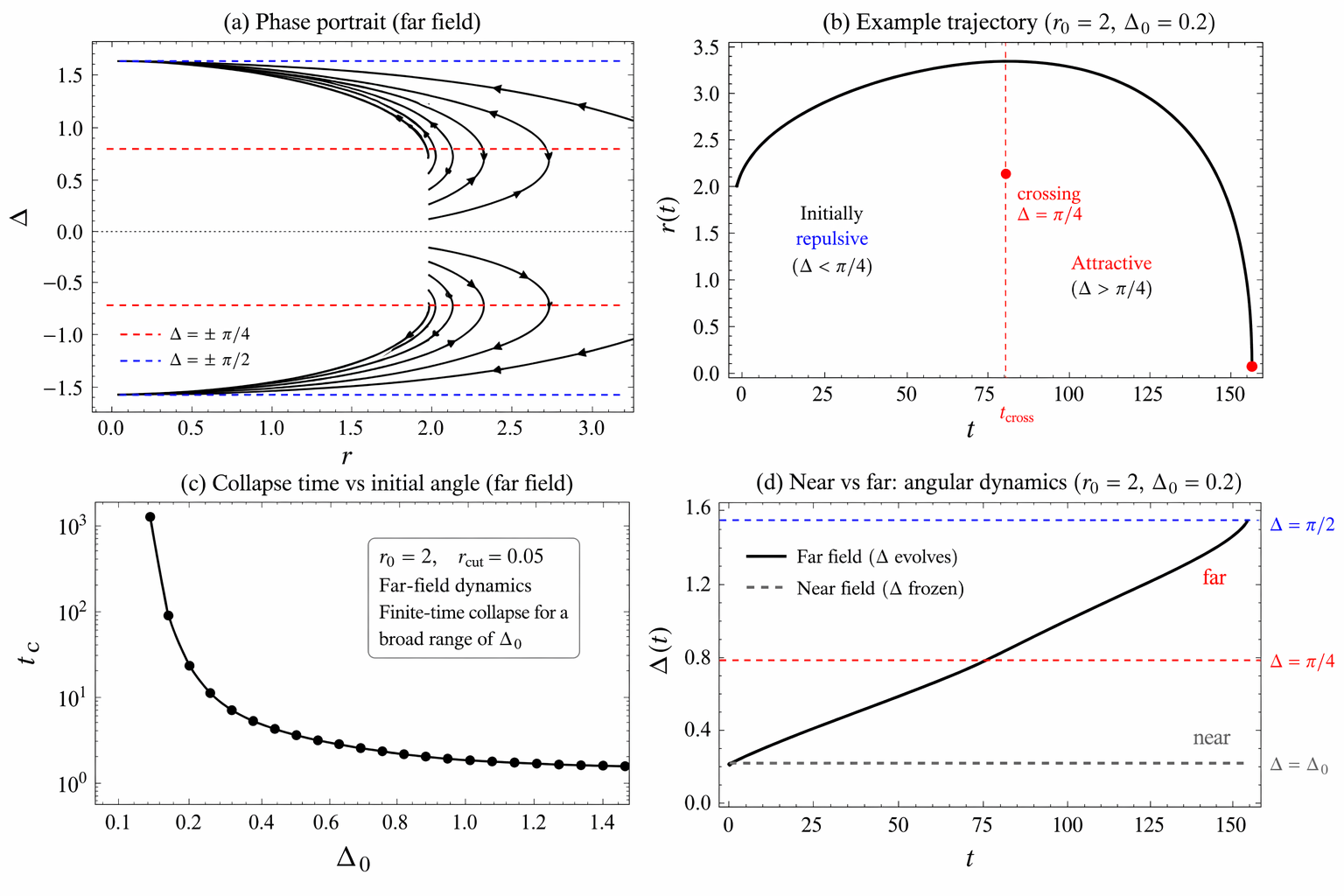}
\caption{
Aggregation mechanism for two quenched dipoles in the screened (far-field) regime. 
(a) Phase portrait in $(r,\Delta)$ showing trajectories for multiple initial conditions; dashed lines denote $\Delta=\pm\pi/4$ (radial nullclines) and $\Delta=\pm\pi/2$. Trajectories bend toward transverse configurations and small separation. 
(b) Representative trajectory starting from an initially repulsive configuration ($\Delta_0<\pi/4$), which crosses into the attractive sector and collapses in finite time. 
(c) Collapse time $t_c$ as a function of initial angle $\Delta_0$, showing aggregation over a broad range of initial conditions. 
(d) Comparison of angular dynamics in near and far fields: in the near field $\Delta$ remains constant, while in the far field it evolves and drives the system across the repulsive–attractive boundary. 
}
\label{fig:far_field_aggregation_story}
\end{figure*}
To elucidate the origin of aggregation in the screened far-field regime, it is useful to examine in detail the reduced two-body dynamics in relative variables. Writing the separation vector as
\[
\mathbf R = \mathbf r_1 - \mathbf r_2 = r\,\hat{\mathbf r}, 
\qquad \hat{\mathbf r}=(\cos\theta,\sin\theta),
\]
and defining the relative orientation mismatch
\[
\Delta = \alpha - \theta,
\]
the quenched co-aligned two-dipole problem reduces to a closed dynamical system for $(r,\Delta)$.

In the unscreened near-field regime, the leading-order dynamics is
\[
\dot r = \frac{\sigma}{2\pi\eta_s r}\cos(2\Delta), 
\qquad 
\dot\Delta = 0.
\]
Thus the angular variable is frozen, and the motion is purely radial. The sign of $\cos(2\Delta)$ partitions phase space into repulsive and attractive sectors separated by the nullclines $\Delta=\pm\pi/4$. However, because $\Delta$ remains constant, trajectories cannot cross between these sectors. As a result, configurations that are initially in the repulsive sector remain non-aggregating for all time, and no collective collapse occurs. The near-field dynamics is therefore effectively one-dimensional despite being represented in a two-dimensional phase space. In contrast, the screened far-field dynamics is governed by
\[
\dot r = \frac{2\sigma}{\pi\zeta_\parallel r^3}\cos(2\Delta), 
\qquad 
\dot\Delta = \frac{2\sigma}{\pi\zeta_\parallel r^4}\sin(2\Delta).
\]
Here radial and angular dynamics are intrinsically coupled. The same nullclines $\Delta=\pm\pi/4$ determine the instantaneous sign of $\dot r$, but these lines are no longer invariant because $\dot\Delta\neq0$ there. Consequently, trajectories generically drift in $\Delta$ and can cross from the initially repulsive sector ($|\Delta|<\pi/4$) into the attractive sector ($|\Delta|>\pi/4$). This angular drift is the key mechanism that enables aggregation. As we found in the main text, a conserved quantity of the far-field system is
\[
\frac{\sin(2\Delta)}{r^2} = \Lambda,
\]
which constrains trajectories to invariant curves in $(r,\Delta)$ space. Along such a curve, as $r\to0$, one must have $\sin(2\Delta)\to0$, which combined with the sign of $\dot\Delta$, selects the asymptotic approach $\Delta\to\pm\pi/2$ for pushers ( $\Delta\to0$ for pullers). Thus, collapse occurs along specific configurations. Importantly, this is different from conventional attractors encountered in dissipative dynamical systems: the flow remains Hamiltonian away from $r=0$. The observed aggregation is therefore a consequence of finite-time approach to a singular manifold rather than dissipative attraction. These features are illustrated in Fig.~\ref{fig:far_field_aggregation_story} for pushers. Panel (a) shows the phase portrait obtained by integrating trajectories from a range of initial conditions. The bending of trajectories toward the lines $\Delta=\pm\pi/2$ and toward small $r$ demonstrates the generic drift into the attractive sector. Panel (b) shows a representative trajectory starting from an initially repulsive configuration ($\Delta_0<\pi/4$). The trajectory first increases in separation, then crosses the nullcline $\Delta=\pi/4$, after which the dynamics becomes attractive and leads to collapse. Panel (c) quantifies the collapse time $t_c$ as a function of the initial angle $\Delta_0$, demonstrating finite-time collapse for a broad range of initial conditions. Panel (d) directly contrasts near- and far-field angular dynamics: in the near field $\Delta(t)$ remains constant, whereas in the far field it evolves monotonically and drives the system towards attractive behavior.

The numerical results were obtained by direct integration of the reduced equations of motion using \textit{Mathematica}. For each initial condition $(r_0,\Delta_0)$, the system was evolved until a cutoff separation $r_{\mathrm{cut}}$ was reached, implemented via an event condition. Phase portraits were constructed by sampling multiple initial angles and plotting the resulting trajectories in the $(r,\Delta)$ plane. Collapse times were extracted from the termination times of the numerical integration. The conservation of the invariant $\sin(2\Delta)/r^2$ was verified numerically to high accuracy, providing an additional consistency check of the integration.

Taken together, the analysis shows that aggregation in the screened membrane is not a consequence of simple radial attraction, but rather emerges from the interplay of angular drift and radial dynamics. The near-field system lacks this mechanism and therefore does not support aggregation, whereas the far-field system generically drives trajectories into the attractive sector, producing robust finite-time collapse. A natural question raised by the phase portraits in Fig.~\ref{fig:far_field_aggregation_story}(a) is whether the far-field dynamics possesses an attractor. 
Although trajectories appear to converge toward the configurations $(r,\Delta)=(0,\pm\pi/2)$ (for pushers), these points do not constitute attractors in the usual dynamical-systems sense. 
The reduced equations remain Hamiltonian away from $r=0$ and admit the conserved quantity $\sin(2\Delta)/r^2$, so phase-space volume is preserved and no dissipative contraction occurs. 
Moreover, the vector field becomes singular as $r\to0$, so the points $(0,\pm\pi/2)$ are not regular fixed points of the flow. 
Instead, the dynamics exhibits finite-time approach to a singular collision manifold. 

\section{Numerical implementation  for near-field many-body quenched dynamics}
\label{app:numerics_near}
\begin{figure}[t]
\centering
\begin{subfigure}[t]{0.48\linewidth}
    \centering
    \includegraphics[width=\linewidth]{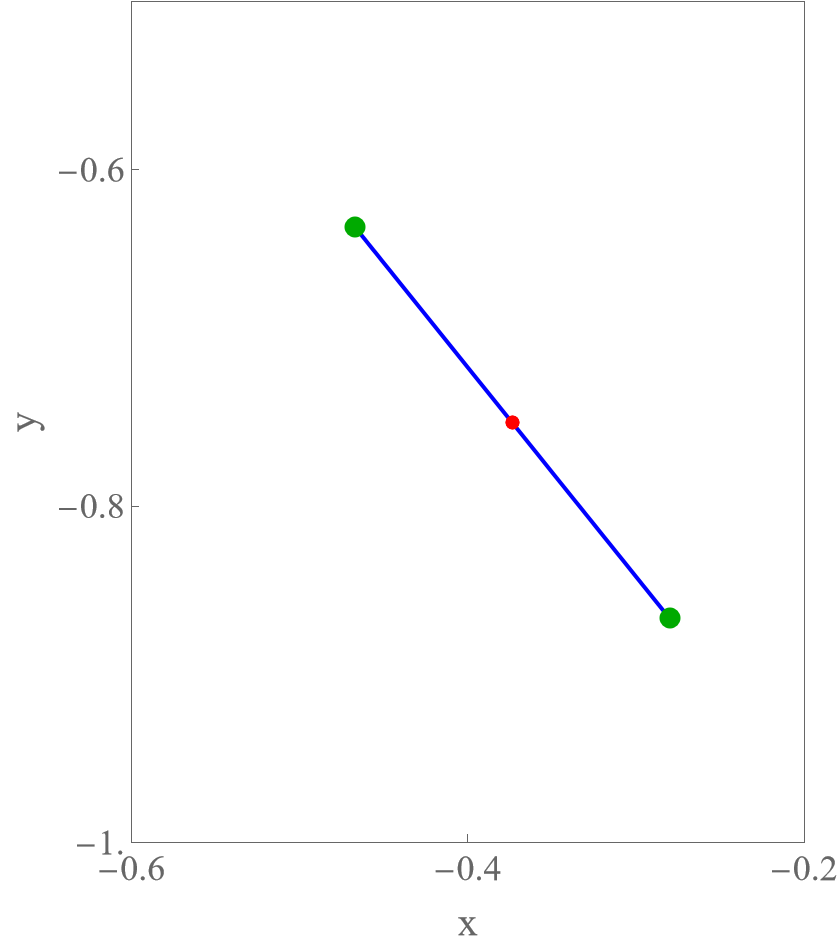}
    \caption{}
\end{subfigure}
\hfill
\begin{subfigure}[t]{0.48\linewidth}
    \centering
    \includegraphics[width=\linewidth]{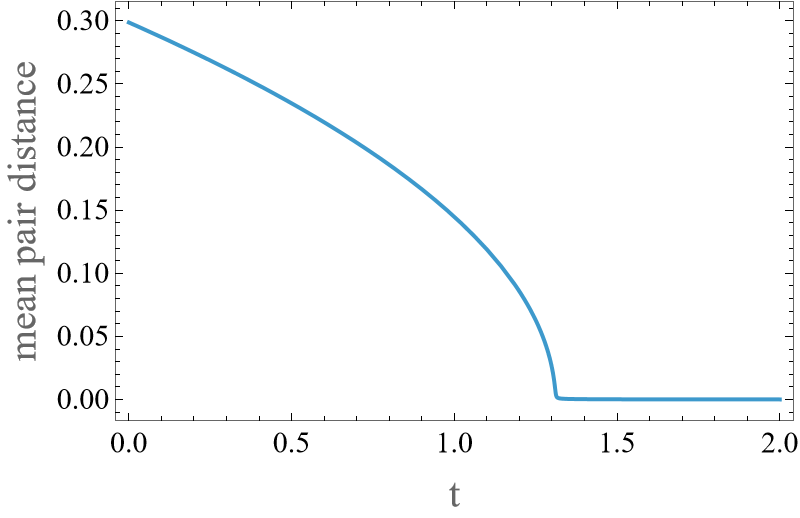}
    \caption{}
\end{subfigure}

\vspace{0.5em}

\begin{subfigure}[t]{0.48\linewidth}
    \centering
    \includegraphics[width=\linewidth]{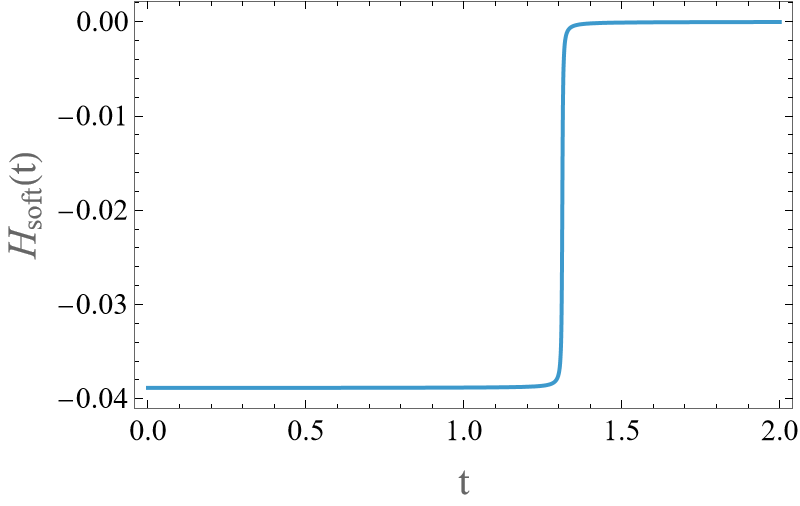}
    \caption{}
\end{subfigure}
\hfill
\begin{subfigure}[t]{0.48\linewidth}
    \centering
    \includegraphics[width=\linewidth]{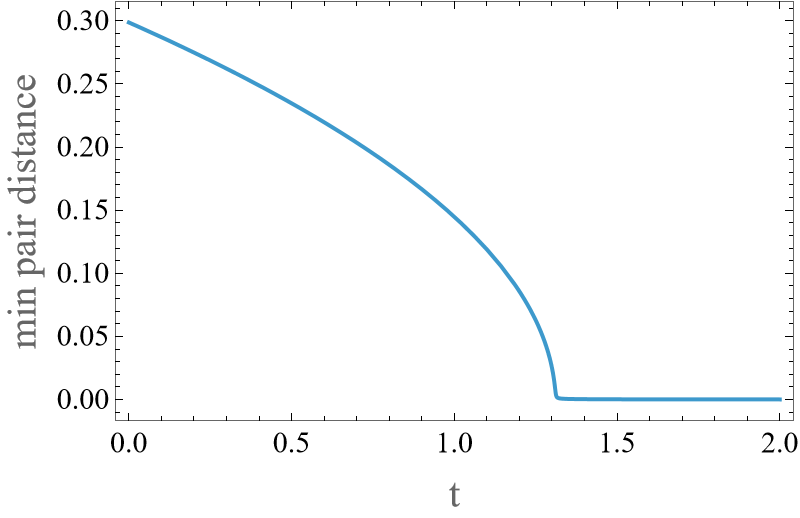}
    \caption{}
\end{subfigure}
\caption{
Two-body numerical illustration of the near-field quenched dynamics (with collapse initial conditions) with soft velocity kernel regularization. Parameters are $N=2$, $\sigma_1=\sigma_2=-1$, $\eta_s=1$, and $\epsilon=0.005$, with initial positions sampled uniformly in the unit disk. (a) Relative trajectory showing a near-collision event; green dots denote initial positions and red dots final positions. (b) Mean pair distance, which for $N=2$ coincides with the pair separation and decreases toward the regularization scale. (c) Time evolution of the softened Hamiltonian  $H_{\rm soft}(t)$, showing a sharp crossover when the separation reaches $r^2\sim\epsilon^2$. (d) Minimum pair distance as a function of time; for $N=2$ this coincides with panel (b) and confirms that the rapid change in $H_{\rm soft}$ occurs precisely at the onset of the regularization-dominated regime.
}
\label{fig:near_two_combined}
\end{figure}
\begin{figure}[t]
\centering
\begin{subfigure}[t]{0.48\linewidth}
    \centering
    \includegraphics[width=\linewidth]{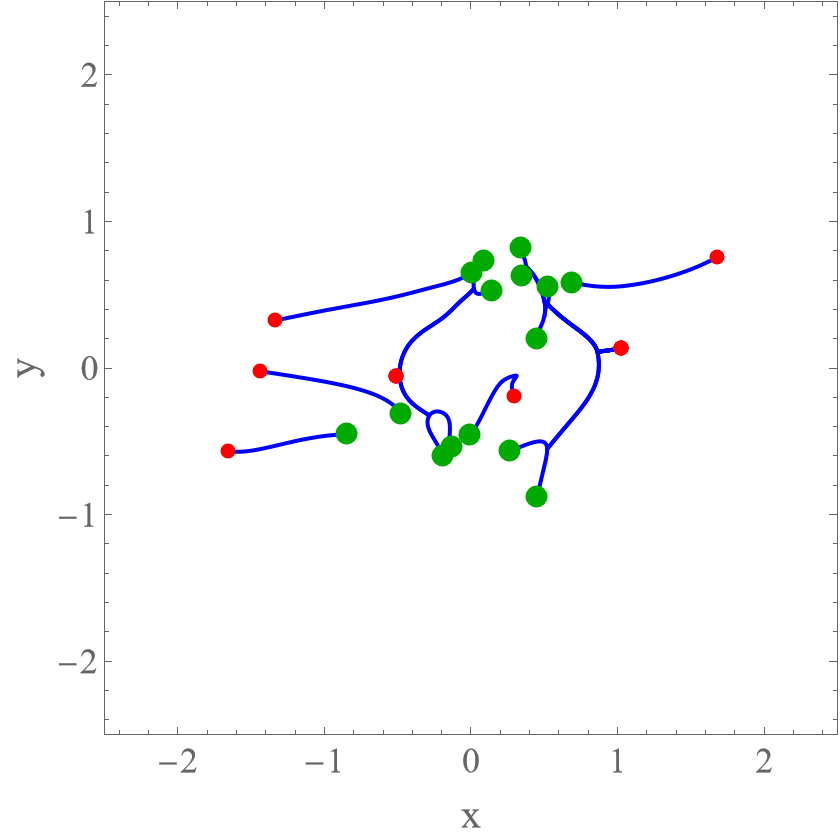}
    \caption{}
\end{subfigure}
\hfill
\begin{subfigure}[t]{0.48\linewidth}
    \centering
    \includegraphics[width=\linewidth]{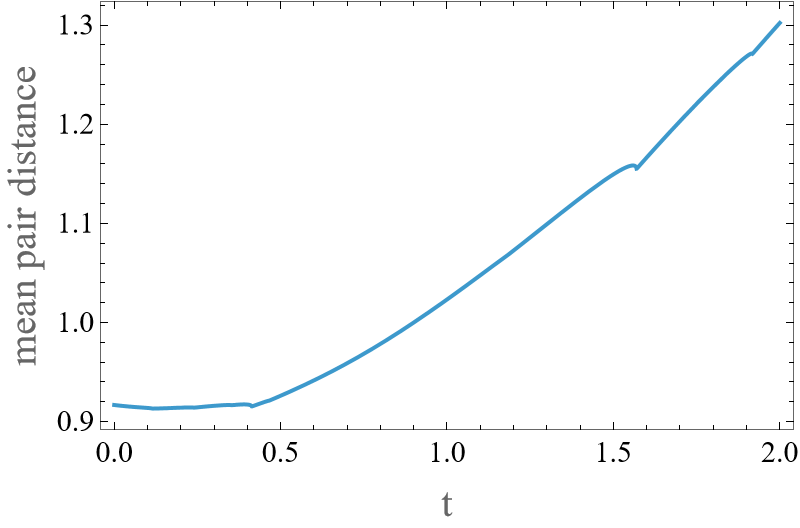}
    \caption{}
\end{subfigure}

\vspace{0.5em}

\begin{subfigure}[t]{0.48\linewidth}
    \centering
    \includegraphics[width=\linewidth]{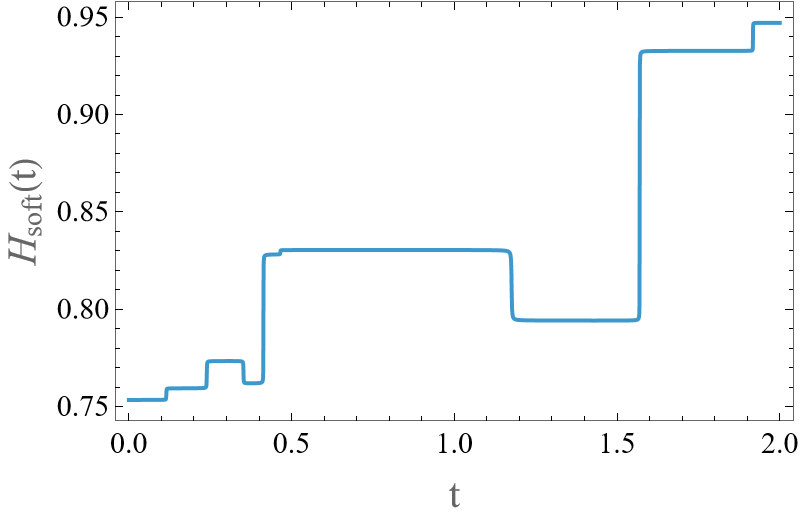}
    \caption{}
\end{subfigure}
\hfill
\begin{subfigure}[t]{0.48\linewidth}
    \centering
    \includegraphics[width=\linewidth]{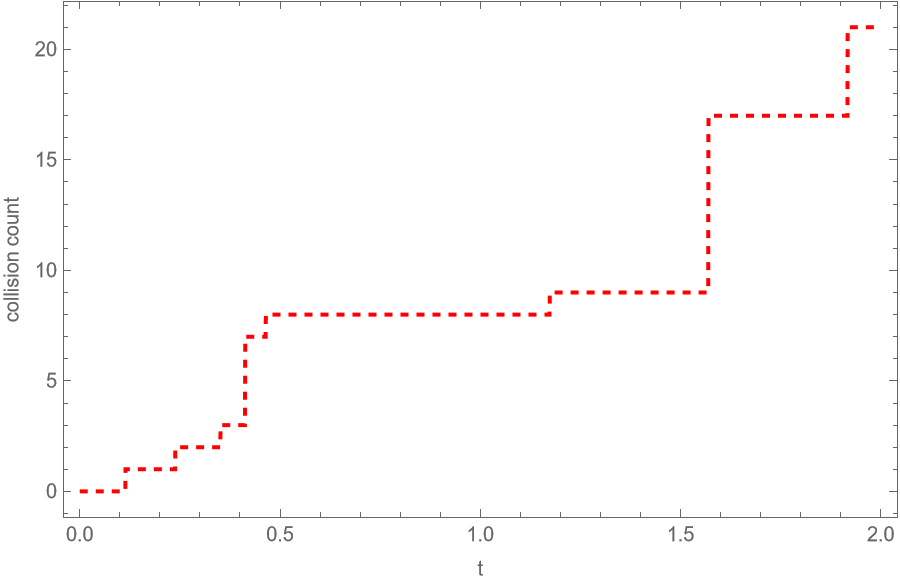}
    \caption{}
\end{subfigure}

\caption{Many-body near-field quenched dynamics with kernel regularization ($N=15$, $\sigma_i=-1$, $\eta_s=1$, $\epsilon=0.005$). (a) Particle trajectories in the plane; green dots denote initial positions and red dots final positions. The motion consists of repeated near-collision events and deflections, without global collapse. (b) Mean pair distance $\langle d_{ij}\rangle(t)$, showing monotonic growth after an initial transient, indicating net dispersal of the ensemble. (c) Softened Hamiltonian  $H_{\rm soft}(t)$ exhibiting a staircase structure with plateaus and sharp jumps. 
(d) Number of near-colliding pairs satisfying $r_{ij}<3\epsilon$ as a function of time. The discrete increases in this quantity are synchronized with the jumps in $H_{\rm soft}(t)$, demonstrating that each jump corresponds to a pair (or set of pairs) entering the regularization zone.
}
\label{fig:near_manybody}
\end{figure}
\begin{figure}[t]
\centering
\begin{subfigure}[t]{0.48\linewidth}
    \centering
    \includegraphics[width=\linewidth]{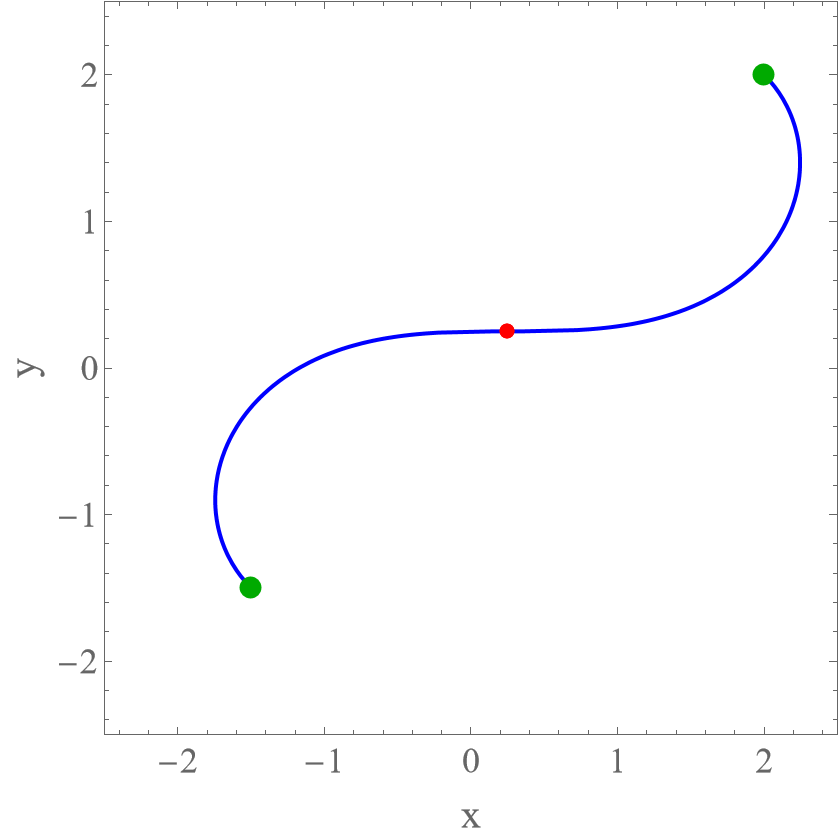}
    \caption{}
\end{subfigure}
\hfill
\begin{subfigure}[t]{0.48\linewidth}
    \centering
    \includegraphics[width=\linewidth]{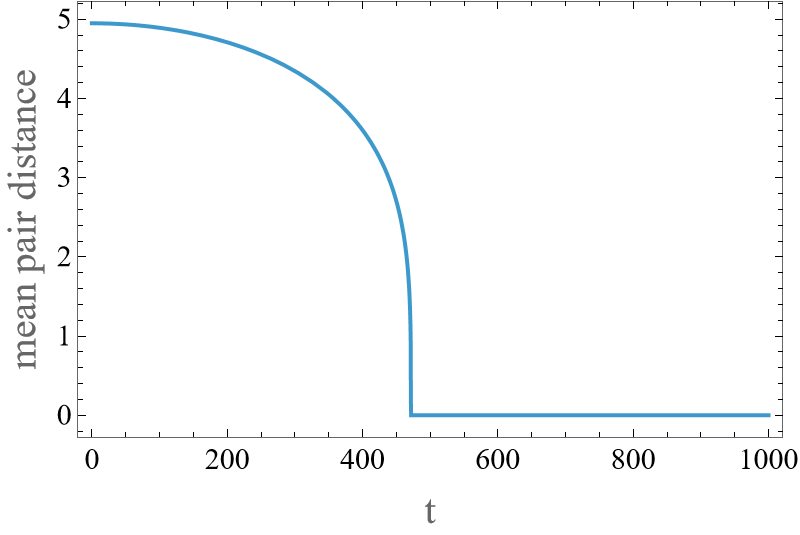}
    \caption{}
\end{subfigure}

\vspace{0.5em}

\begin{subfigure}[t]{0.48\linewidth}
    \centering
    \includegraphics[width=\linewidth]{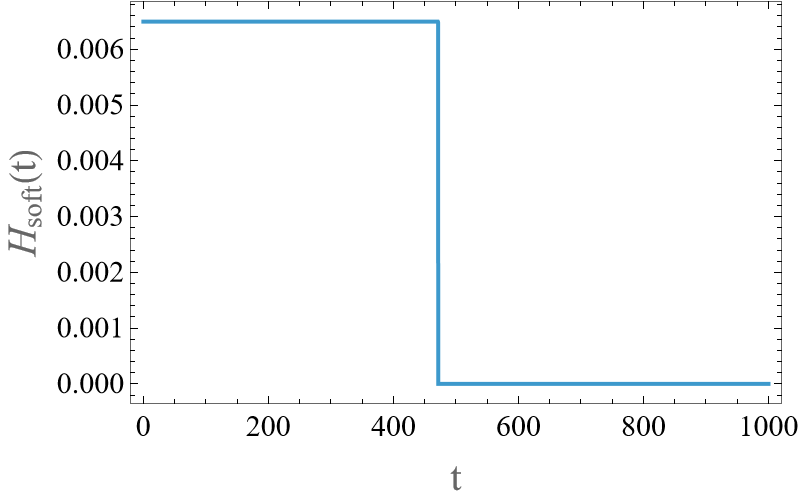}
    \caption{}
\end{subfigure}
\hfill
\begin{subfigure}[t]{0.48\linewidth}
    \centering
    \includegraphics[width=\linewidth]{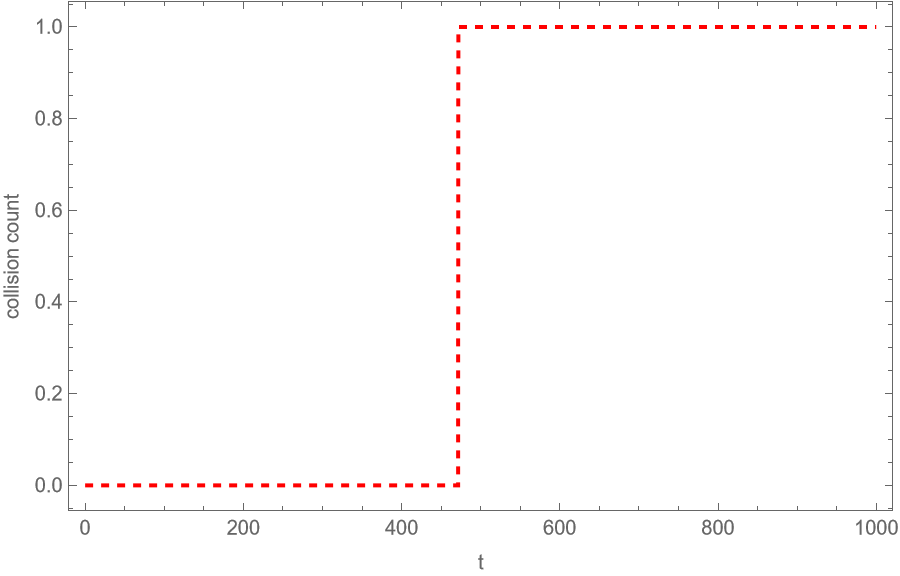}
    \caption{}
\end{subfigure}
\caption{Two-body numerical illustration of far-field quenched dynamics with kernel regularization. Parameters: $N=2$, $\sigma_1=\sigma_2=-1$, $\zeta_\parallel=1$, $\epsilon=0.005$, with initial positions $\mathbf r_1(0)=(-1.5,-1.5)$ and $\mathbf r_2(0)=(2,2)$. 
(a) Particle trajectories in the plane; green dots denote initial positions and the red dot marks the common near-collision location. 
(b) Mean pair distance (equal to the interparticle separation for $N=2$). The extended initial shoulder reflects slow angular drift away from the radial nullcline, followed by rapid collapse. 
(c) Time evolution of the softened Hamiltonian $H_{\rm soft}(t)$, nearly constant outside the regularization zone and dropping sharply when $r\sim\epsilon$. 
(d) Collision-count (number of pairs with $r_{ij}<3\epsilon$); for $N=2$ it switches from $0$ to $1$ when the pair enters the softening zone, confirming that the abrupt change in $H_{\rm soft}(t)$ originates from regularized short-distance dynamics.}
\label{fig:far_two_combined}
\end{figure}
\begin{figure}[t]
\centering
\begin{subfigure}[t]{0.48\linewidth}
    \centering
    \includegraphics[width=\linewidth]{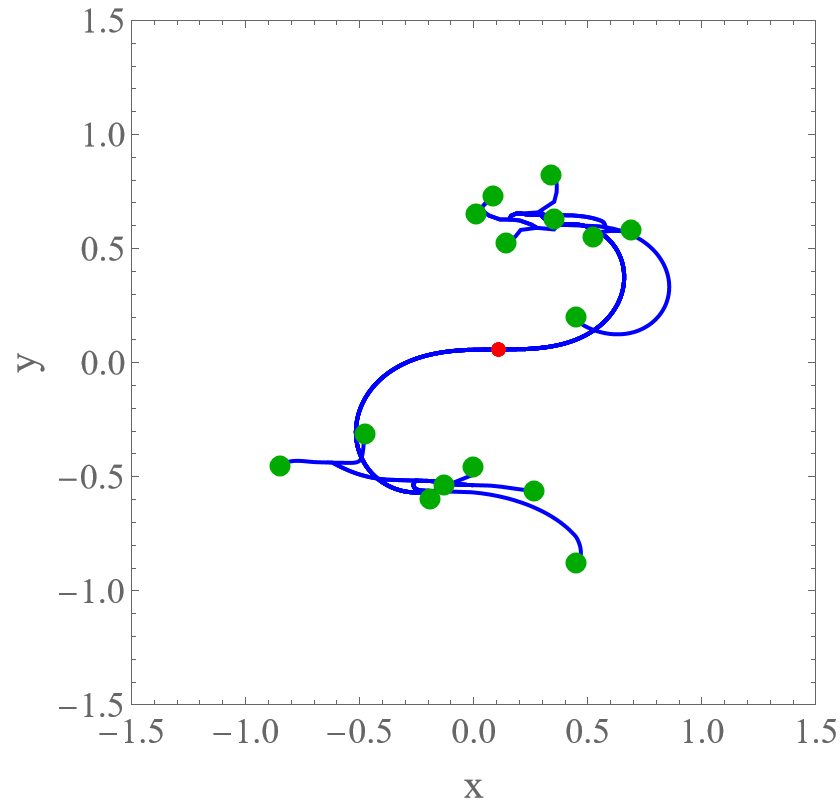}
    \caption{}
\end{subfigure}
\hfill
\begin{subfigure}[t]{0.48\linewidth}
    \centering
    \includegraphics[width=\linewidth]{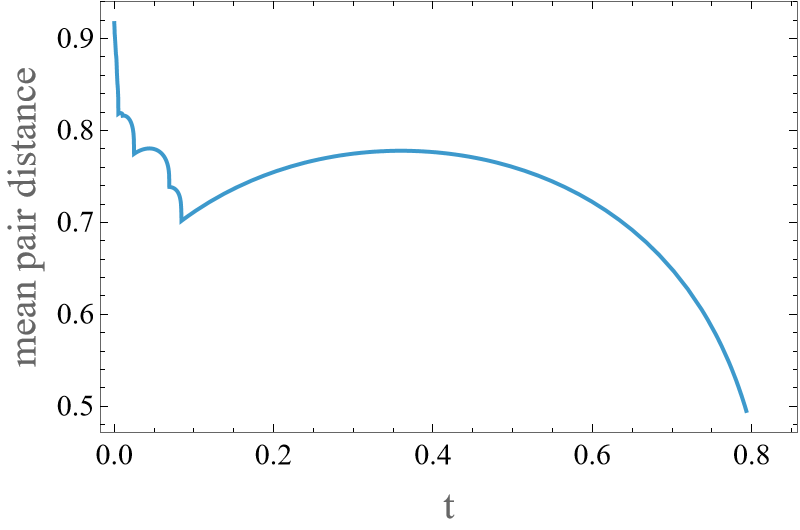}
    \caption{}
\end{subfigure}

\vspace{0.5em}

\begin{subfigure}[t]{0.48\linewidth}
    \centering
    \includegraphics[width=\linewidth]{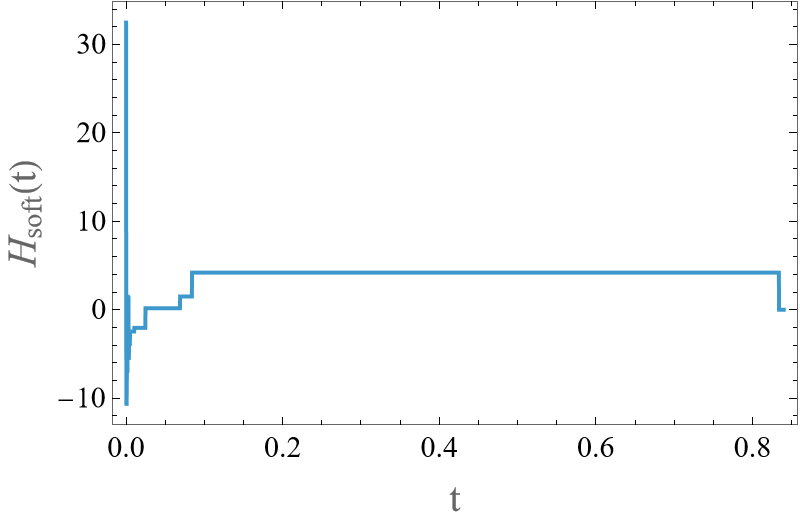}
    \caption{}
\end{subfigure}
\hfill
\begin{subfigure}[t]{0.48\linewidth}
    \centering
    \includegraphics[width=\linewidth]{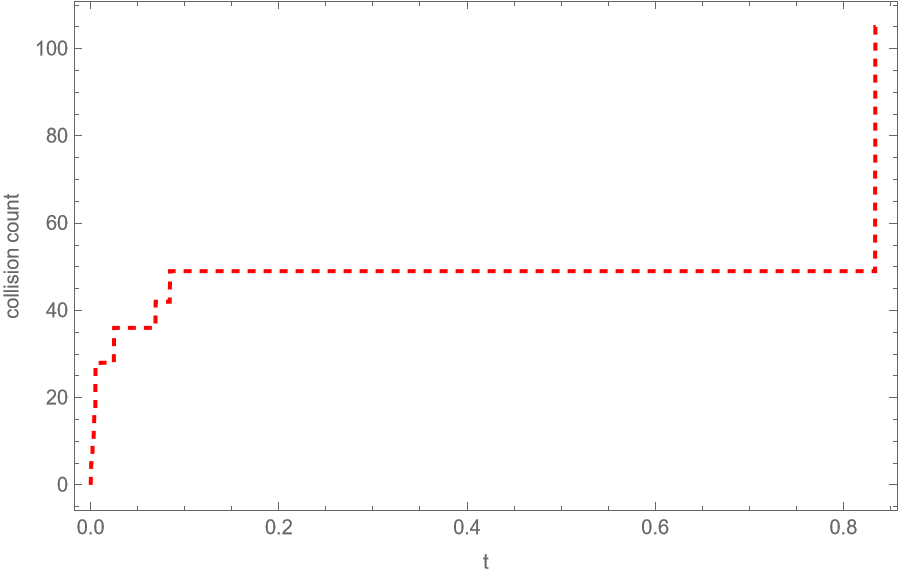}
    \caption{}
\end{subfigure}
\caption{Many-body numerical illustration of far-field quenched dynamics with  kernel regularization. Parameters: $N=15$, $\sigma_i=-1$, $\zeta_\parallel=1$, $\epsilon=0.005$, with initial positions sampled uniformly in the unit disk. (a) Particle trajectories; green dots mark initial positions and the red dot the common softened endpoint. Screened far-field interactions rapidly collect the ensemble into a compact region, followed by condensation into a single softened cluster. (b) Mean pair distance $\langle d_{ij}\rangle(t)$: an initial rapid decrease signals cluster formation, a mild increase reflects internal reorganization, and a final sharp drop marks collapse into the softened core. (c) Softened Hamiltonian $H_{\rm soft}(t)$: large early excursions indicate rapid rearrangement, a later plateau corresponds to two long-lived compact aggregates, and the final decay to zero occurs when the clusters merge (d) Collision-count  (pairs with $r_{ij}<3\epsilon$): a rapid rise to an intermediate plateau indicates early formation of close pairs, while the final jump to $\binom{15}{2}=105$ signifies complete clustering.}
\label{fig:far_manybody_combined}
\end{figure}
In this Appendix we describe the numerical implementation of the near-field quenched dipole dynamics discussed in Sec.~\ref{sec:hamiltonian}, and analyze in detail the behavior of the diagnostics used to characterize the evolution. We consider $N$ co-aligned dipoles with fixed orientation $\alpha_i=\alpha=0$, evolving under the near-field Hamiltonian (in cartesian variables)
\begin{equation}
H_{\rm near}
=
\frac{1}{4\pi\eta_s}
\sum_{i<j}
\sigma_i\sigma_j
\frac{\Delta x_{ij}\Delta y_{ij}}{r_{ij}^2},
\end{equation}
where $\Delta x_{ij}=x_i-x_j$, $\Delta y_{ij}=y_i-y_j$, and $r_{ij}^2=\Delta x_{ij}^2+\Delta y_{ij}^2$. The corresponding Hamiltonian equations,
\begin{equation}
\dot{x}_i = \frac{1}{\sigma_i}\frac{\partial H}{\partial y_i},
\qquad
\dot{y}_i = -\frac{1}{\sigma_i}\frac{\partial H}{\partial x_i},
\end{equation}
generate pairwise velocities with a $1/r^4$ singularity. To enable stable numerical integration we introduce a soft regulator by replacing
\begin{equation}
r_{ij}^4 \longrightarrow (r_{ij}^2+\epsilon^2)^2
\end{equation}
only in the velocity kernel, while keeping the algebraic structure of the numerators unchanged. The resulting pairwise contributions to the velocity of particle $i$ due to particle $j$ are
\begin{align}
v_x^{(ij)}
&=
\sigma_j\,\frac{1}{4\pi\eta_s}
\frac{\Delta x_{ij}(-\Delta x_{ij}^2+\Delta y_{ij}^2)}{(r_{ij}^2+\epsilon^2)^2},
\\
v_y^{(ij)}
&=
-\sigma_j\,\frac{1}{4\pi\eta_s}
\frac{\Delta y_{ij}(\Delta x_{ij}^2-\Delta y_{ij}^2)}{(r_{ij}^2+\epsilon^2)^2}.
\end{align}
This regularization removes the singularity but modifies the dynamics: the regularized evolution preserves the Hamiltonian structure between collisions, while altering the behavior during collisions and close encounters; its variation instead provides a sensitive indicator of the onset of the regularization-dominated regime. Initial particle positions are sampled uniformly within the unit disk according to
\begin{equation}
r=\sqrt{\xi},\qquad \theta=2\pi\zeta,
\end{equation}
with $\xi,\zeta\in[0,1]$ uniformly distributed random variables. We take identical dipoles with $\sigma_i=-1$. The equations of motion are integrated numerically with adaptive step-size control. To characterize the evolution we monitor the mean pair distance
\begin{equation}
\langle d_{ij}\rangle(t)
=
\frac{2}{N(N-1)}
\sum_{i<j} r_{ij}(t),
\end{equation}
the minimum pair distance
\begin{equation}
d_{\min}(t)=\min_{i<j} r_{ij}(t),
\end{equation}
and the softened Hamiltonian 
\begin{equation}
H_{\rm soft}(t)
=
\frac{1}{4\pi\eta_s}
\sum_{i<j}
\sigma_i\sigma_j
\frac{\Delta x_{ij}\Delta y_{ij}}{r_{ij}^2+\epsilon^2},
\label{eq:Hsoft_def_app}
\end{equation}
which reduces to the unregularized Hamiltonian as $\epsilon\to0$. Because the equations of motion are generated by the softened velocities, the quantity $H_{\rm soft}$ is not exactly conserved near collisions;  The interpretation of evolution of $H_{\rm soft}$ is especially simple in the two-body case worked out in detail below. For $N=2$, introducing the relative coordinate $\mathbf R=(\Delta x,\Delta y)$ with $u=r^2=\Delta x^2+\Delta y^2$, one finds that the relative angle $\theta=\arg(\mathbf R)$ remains fixed in the near-field regime, and the dynamics closes to
\begin{equation}
\dot u
=
-c\,\frac{u^2}{(u+\epsilon^2)^2},
\qquad
c=4\sigma\,\frac{1}{4\pi\eta_s}\cos 2\theta_0,
\label{eq:u_eqn_app}
\end{equation}
where $\theta_0$ is the initial angle of the line of centers. In the same two-body problem, Eq.~\eqref{eq:Hsoft_def_app} reduces to
\begin{equation}
H_{\rm soft}(t)
=
\frac{1}{8\pi\eta_s}\sin 2\theta_0\,
\frac{u(t)}{u(t)+\epsilon^2}.
\label{eq:Hsoft_two_app}
\end{equation}
For $u\gg\epsilon^2$, the ratio $u/(u+\epsilon^2)$ is close to unity, so $H_{\rm soft}$ remains approximately constant and reproduces the unregularized Hamiltonian behavior. Once the relative separation reaches the softening scale, $u\sim\epsilon^2$, Eq.~\eqref{eq:Hsoft_two_app} predicts a rapid crossover of $H_{\rm soft}$ toward zero. Since the temporal width of this crossover is of order $\Delta t\sim \epsilon^2/c$, it is extremely narrow when $\epsilon\ll1$, and therefore appears in the numerics as a sharp jump. The apparent discontinuity is thus not a genuine non-smoothness of the underlying differential equation, but rather a manifestation of the small regularization scale together with the use of $H_{\rm soft}$ as a diagnostic quantity.

A representative numerical example illustrating this mechanism is shown in Fig.~\ref{fig:near_two_combined}. The calculation uses $N=2$ identical dipoles with $\sigma_1=\sigma_2=-1$, $\eta_s=1$, and $\epsilon=0.005$, with initial positions sampled from the unit disk according to the procedure described above. The trajectory plot in Fig.~\ref{fig:near_two_combined}(a) shows that the motion remains collinear and leads to a near-collision event, as expected from the reduced two-body structure. The corresponding pair distance in Fig.~\ref{fig:near_two_combined}(b) decreases at early times, consistent with the unregularized near-zone collapse law, and then rapidly saturates once the separation falls to the softening scale. The same event is reflected in the softened Hamiltonian shown in Fig.~\ref{fig:near_two_combined}(c): $H_{\rm soft}(t)$ stays nearly constant over most of the trajectory and then undergoes a sharp crossover when the pair enters the regularization zone. Finally, Fig.~\ref{fig:near_two_combined}(d) shows the minimum pair distance, which for $N=2$ coincides with the mean pair distance and pinpoints the time at which the crossover occurs. The comparison of panels (b)-(d) makes clear that the sudden change in $H_{\rm soft}$ is synchronized with the approach of the pair to the scale $r^2\sim \epsilon^2$. These results therefore confirm that the observed jumps in $H_{\rm soft}$ originate from the interplay of near-field collapse dynamics with the finite softening length, rather than from any intrinsic discontinuity of the underlying flow.\\
We now extend the above analysis to a genuinely many-body system and demonstrate that the two-body crossover mechanism persists as a sequence of localized events. We consider $N=15$ identical co-aligned dipoles with $\sigma_i=-1$ (pullers), $\eta_s=1$, and $\epsilon=0.005$, with initial positions sampled uniformly in the unit disk as described above. The equations of motion are integrated up to $t=2$ using adaptive time stepping.

A representative realization is shown in Fig.~\ref{fig:near_manybody}. The trajectory plot in panel (a) reveals that the dynamics is dominated by a sequence of close encounters between particles. While several dipoles undergo near-collision events and form transient clusters, the motion is not globally collapsing: particles repeatedly approach, deflect, and reorganize. 

The global behavior is quantified in panel (b), which shows the mean pair distance $\langle d_{ij}\rangle(t)$. After a short initial transient, the mean separation increases monotonically in time, indicating that the system as a whole expands rather than collapses. This demonstrates that, despite strong pairwise attraction in certain angular sectors, the unscreened $1/r$ interaction does not produce sustained many-body aggregation. Instead, the competing repulsive sectors lead to a net dispersive effect at the ensemble level.

The microscopic origin of this behavior is captured by the softened Hamiltonian $H_{\rm soft}(t)$, shown in panel (c). In contrast to the smooth evolution of geometric observables, $H_{\rm soft}(t)$ exhibits a pronounced staircase structure consisting of extended plateaus separated by sharp jumps. This structure is a direct many-body generalization of the two-body crossover discussed above. When all pair separations satisfy $r_{ij}^2\gg\epsilon^2$, each term in Eq.~\eqref{eq:Hsoft_def_app} is close to its unregularized value and $H_{\rm soft}(t)$ remains approximately constant. However, when one or more pairs approach the softening scale $r_{ij}\sim\epsilon$, the corresponding contributions are rapidly suppressed, producing a finite change in $H_{\rm soft}$. The magnitude and sign of each jump depend on the angular factor $\Delta x_{ij}\Delta y_{ij}$, so that successive events can either increase or decrease $H_{\rm soft}$.

To make this connection explicit, panel (d) shows the time evolution of the number of near-colliding pairs, defined as the number of pairs satisfying $r_{ij}(t)<3\epsilon$. This quantity increases in discrete steps, each corresponding to the entry of additional particle pairs into the regularization zone. A direct comparison between panels (c) and (d) shows that the jumps in $H_{\rm soft}(t)$ are synchronized with these increments: each step in the collision count coincides with a transition between plateaus in $H_{\rm soft}$. This establishes that the staircase structure of $H_{\rm soft}(t)$ is the cumulative signature of localized near-collision events.

Taken together, these results demonstrate that the many-body near-field dynamics proceeds through a sequence of effectively two-body processes, each governed by the same mechanism identified in the $N=2$ problem. While these localized interactions produce transient clustering and rapid rearrangements, they do not lead to global collapse. Instead, the system evolves through repeated near-collision events that redistribute particles and ultimately drive an overall expansion of the ensemble. This provides a clear dynamical explanation for the suppression of aggregation in the unscreened near-field regime.

\section{Numerical implementation for far-field quenched dynamics}
\label{app:numerics_far}
In this Appendix we describe the numerical implementation of the screened far-field quenched dipole dynamics discussed in Sec.~\ref{sec:hamiltonian}, and illustrate its behavior in a representative two-body test. We restrict throughout to the co-aligned sector $\alpha_i=\alpha=0$, for which the far-field Hamiltonian takes the Cartesian form
\begin{equation}
H_{\rm far}
=
\frac{1}{\pi\zeta_\parallel}
\sum_{i<j}
\sigma_i\sigma_j
\frac{\Delta x_{ij}\Delta y_{ij}}{r_{ij}^4},
\label{eq:Hfar_app}
\end{equation}
where $\Delta x_{ij}=x_i-x_j$, $\Delta y_{ij}=y_i-y_j$, and $r_{ij}^2=\Delta x_{ij}^2+\Delta y_{ij}^2$. The equations of motion are written in Hamiltonian form as
\begin{equation}
\dot{x}_i = \frac{1}{\sigma_i}\frac{\partial H}{\partial y_i},
\qquad
\dot{y}_i = -\frac{1}{\sigma_i}\frac{\partial H}{\partial x_i},
\label{eq:far_ham_app}
\end{equation}
and yield pairwise velocities with a $1/r^6$ singularity in Cartesian coordinates. Differentiating Eq.~\eqref{eq:Hfar_app} gives the pair contribution of particle $j$ to the velocity of particle $i$,
\begin{align}
v_x^{(ij)}
&=
\sigma_j\,\frac{1}{\pi\zeta_\parallel}
\frac{\Delta x_{ij}\left(\Delta x_{ij}^2-3\Delta y_{ij}^2\right)}{r_{ij}^6},
\\
v_y^{(ij)}
&=
\sigma_j\,\frac{1}{\pi\zeta_\parallel}
\frac{\Delta y_{ij}\left(3\Delta x_{ij}^2-\Delta y_{ij}^2\right)}{r_{ij}^6}.
\label{eq:far_pairvel_app}
\end{align}
As in the near-zone numerics, direct integration of these equations becomes unstable during close encounters because of the singular short-distance kernel. To continue trajectories through such events we introduce a softening at the level of the velocity field only, replacing
\begin{equation}
r_{ij}^6 \longrightarrow (r_{ij}^2+\epsilon^2)^3,
\end{equation}
while leaving the algebraic numerators unchanged. The softened pairwise velocities used in the numerics are therefore
\begin{align}
v_x^{(ij)}
&=
\sigma_j\,\frac{1}{\pi\zeta_\parallel}
\frac{\Delta x_{ij}\left(\Delta x_{ij}^2-3\Delta y_{ij}^2\right)}{(r_{ij}^2+\epsilon^2)^3},
\\
v_y^{(ij)}
&=
\sigma_j\,\frac{1}{\pi\zeta_\parallel}
\frac{\Delta y_{ij}\left(3\Delta x_{ij}^2-\Delta y_{ij}^2\right)}{(r_{ij}^2+\epsilon^2)^3}.
\label{eq:far_softpairvel_app}
\end{align}
This regularization behaves similarly to the near-field case: conservation holds between collsions. To monitor the evolution we use the same geometric observables as in the near-field case. The mean pair distance is
\begin{equation}
\langle d_{ij}\rangle(t)
=
\frac{2}{N(N-1)}
\sum_{i<j}r_{ij}(t),
\end{equation}
and the minimum pair distance is
\begin{equation}
d_{\min}(t)=\min_{i<j}r_{ij}(t).
\end{equation}
The natural far-field softened Hamiltonian is obtained from Eq.~\eqref{eq:Hfar_app} by replacing $r_{ij}^4$ by $(r_{ij}^2+\epsilon^2)^2$,
\begin{equation}
H_{\rm soft}(t)
=
\frac{1}{\pi\zeta_\parallel}
\sum_{i<j}
\sigma_i\sigma_j
\frac{\Delta x_{ij}\Delta y_{ij}}{(r_{ij}^2+\epsilon^2)^2}.
\label{eq:Hsoft_far_app}
\end{equation}
In the limit $r_{ij}^2\gg\epsilon^2$, this quantity approaches the unregularized Hamiltonian and is constant, whereas once a pair enters the regularization zone $r_{ij}\sim\epsilon$, the corresponding contribution is rapidly suppressed; its variation instead provides a sensitive indicator of the onset of the regularization-dominated regime. To localize such events directly in configuration space, we also monitor the collision-count  defined as the number of pairs satisfying $r_{ij}(t)<3\epsilon$.

A representative two-body calculation is shown in Fig.~\ref{fig:far_two_combined}. We consider $N=2$ identical pullers with $\sigma_1=\sigma_2=-1$, $\zeta_\parallel=1$, and $\epsilon=0.005$, integrated up to $t_{\max}=1000$. The initial positions are chosen as
\begin{equation}
\mathbf r_1(0)=(-1.5,-1.5),
\qquad
\mathbf r_2(0)=(2,2),
\label{eq:far_IC_app}
\end{equation}
so that the initial relative vector lies along the diagonal $y=x$. In the co-aligned sector $\alpha=0$, this places the system on the critical far-field branch for which the instantaneous radial drift vanishes while the angular drift remains nonzero. The resulting trajectory in Fig.~\ref{fig:far_two_combined}(a) is therefore not a straight inward approach: the pair first rotates away from the critical direction and only then develops radial attraction. This produces the characteristic curved,  paths seen in the figure, with the center of mass remaining fixed throughout.

The corresponding mean pair distance is shown in Fig.~\ref{fig:far_two_combined}(b). Since $N=2$, this quantity coincides exactly with the actual interparticle separation. The separation decreases slowly over an extended time interval and then drops rapidly near the end of the evolution. This delayed collapse is a direct consequence of the critical initial orientation. Because the initial radial velocity vanishes, the dynamics begins with primarily angular reorganization. Only after the line of centers has rotated away from the nullcline does a substantial inward radial drift develop, leading to accelerated collapse. Thus, the long shoulder region visible in Fig.~\ref{fig:far_two_combined}(b) is the expected geometric signature of starting on the radial nullcline of the screened two-body system.

The same event is captured by the softened Hamiltonian diagnostic shown in Fig.~\ref{fig:far_two_combined}(c). For most of the evolution, $H_{\rm soft}(t)$ remains nearly constant, reflecting the fact that the pair is well outside the regularization zone and the softened kernel is effectively indistinguishable from the unregularized one. Once the separation reaches $r\sim\epsilon$, however, the far-field kernel becomes dominated by the softening scale and the contribution to Eq.~\eqref{eq:Hsoft_far_app} is rapidly driven to zero. The apparent discontinuity of $H_{\rm soft}(t)$ is therefore not a physical non-smoothness of the underlying flow; rather, it is a diagnostic signature of entry into the softening zone, analogous to the sharp crossover discussed in the near-field two-body problem, but now for the screened $1/r^3$ interaction.

This interpretation is confirmed directly by the collision-count in Fig.~\ref{fig:far_two_combined}(d). For $N=2$, the collision count can take only the values $0$ and $1$. It remains zero throughout the slow approach stage and jumps to unity precisely when the particles enter the threshold region $r<3\epsilon$. The timing of this jump coincides with both the collapse of the mean pair distance and the abrupt suppression of $H_{\rm soft}(t)$. The four panels of Fig.~\ref{fig:far_two_combined} therefore provide a consistent numerical realization of the screened far-field two-body dynamics: an initially critical configuration first undergoes angular drift, then develops radial attraction, and finally enters the regularization-dominated regime, where the softened Hamiltonian collapses and the collision counter switches on.\\
A representative many-body realization is shown in Fig.~\ref{fig:far_manybody_combined}, where we consider $N=15$ identical pullers with $\sigma_i=-1$, $\zeta_\parallel=1$, and $\epsilon=0.005$, initialized at random positions sampled uniformly from the unit disk and evolved up to $t_{\max}=0.84$. In contrast to the two-body example of Fig.~\ref{fig:far_two_combined}, the many-body system displays a richer multistage aggregation process in which rapid early clustering is followed by a prolonged compact, but not yet fully collapsed, state and only later by complete condensation into the softened core.

The trajectory plot in Fig.~\ref{fig:far_manybody_combined}(a) shows that all particles are drawn toward a common central region, confirming the strongly attractive character of the screened far-field Hamiltonian in the co-aligned limit. The trajectories are not straight radial lines; instead they exhibit curved approach, looping, and local rearrangements as different particles are incorporated into the growing aggregate. Nevertheless, the global trend is unambiguous: the ensemble rapidly contracts from its initially diffuse configuration into a compact cluster and ultimately condenses into a common softened endpoint. This behavior is qualitatively different from the near-field many-body dynamics discussed above, where close encounters do not lead to irreversible global aggregation.

The corresponding mean pair distance $\langle d_{ij}\rangle(t)$ is shown in Fig.~\ref{fig:far_manybody_combined}(b). The evolution has three distinct stages. First, there is a sharp early drop from the initial value, indicating prompt formation of two compact clusters. Second, after this rapid contraction, the mean pair distance increases slightly over an extended interval. This increase does not signal global dispersal. Rather, it reflects internal reorganization within the already aggregated states. Third, after this restructuring stage, the mean pair distance decreases again and finally drops sharply near the end of the run, marking the onset of complete collapse into the softened core. Thus, the nonmonotonic behavior of $\langle d_{ij}\rangle(t)$ is the signature of a cluster-then-collapse scenario rather than a simple direct contraction.

This interpretation is strongly supported by the collision-count  in Fig.~\ref{fig:far_manybody_combined}(d), defined as the number of pairs satisfying $r_{ij}(t)<3\epsilon$. For $N=15$, the total number of distinct pairs is $\binom{15}{2}=105$. The collision count rises very rapidly at early times to a plateau near half of this maximal value, demonstrating that a substantial fraction of the system has already entered the regularization-dominated near-contact regime shortly after the evolution begins. Importantly, however, the count does not immediately reach $105$. The long intermediate plateau shows that the system has formed a two compact metastable aggregates. Only near the end of the run does the collision count jump to its maximal value, indicating that every pair has entered the softened core and that the entire ensemble has condensed into a single cluster.
The softened Hamiltonian $H_{\rm soft}(t)$ shown in Fig.~\ref{fig:far_manybody_combined}(c) tells the same story. At very early times it exhibits large, rapidly varying excursions of both signs. These fluctuations arise because the total  is a sum of many pair contributions,
\[
\propto
\frac{\Delta x_{ij}\Delta y_{ij}}{(r_{ij}^2+\epsilon^2)^2},
\]
whose magnitudes and signs change abruptly as numerous pairs simultaneously approach one another and reorganize geometrically. The strong oscillations are therefore not numerical artifacts; they are the natural many-body counterpart of the sharp crossover already identified in the two-body problem, now occurring repeatedly and collectively during the rapid cluster-formation stage. After this initial transient, $H_{\rm soft}(t)$ settles to a broad plateau, consistent with the existence of two compact aggregates whose overall pair geometry changes only slowly. Finally, when the full ensemble enters the softened core, $H_{\rm soft}(t)$ is driven rapidly toward zero, reflecting the suppression of all pair contributions by the regularized denominator once $r_{ij}\sim\epsilon$ for essentially every pair.  The many-body far-field numerics therefore complements the two-body example of Fig.~\ref{fig:far_two_combined} by showing how the same screened attraction that leads to pair collapse is amplified collectively into a many-body aggregation.

\section{Simulations using soft harmonic repulsion}
\label{supfig}
\begin{figure*}[t]
  \centering

  \begin{subfigure}[t]{0.35\textwidth}
    \centering
    \includegraphics[width=\linewidth]{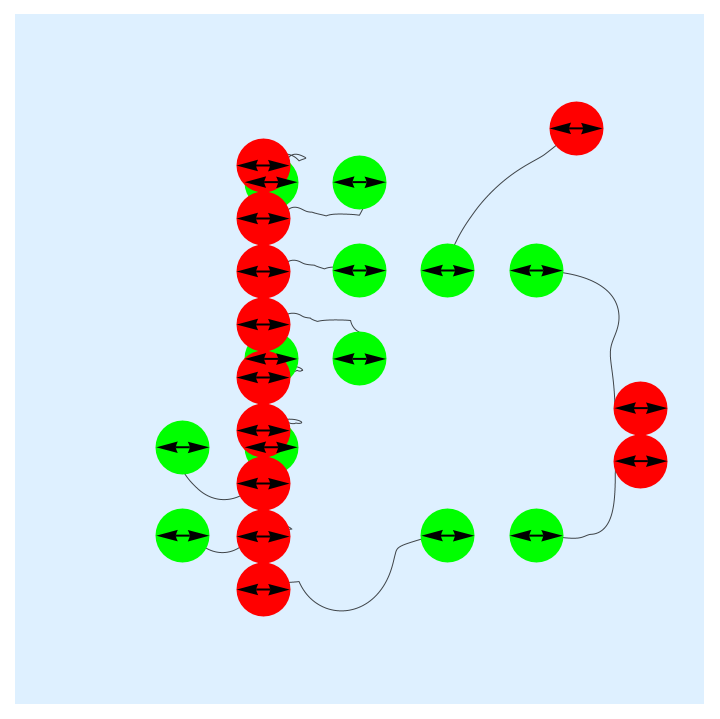}
    \caption{Pushers (far zone)}
  \end{subfigure}
  \hfill
  \begin{subfigure}[t]{0.35\textwidth}
    \centering
    \includegraphics[width=\linewidth]{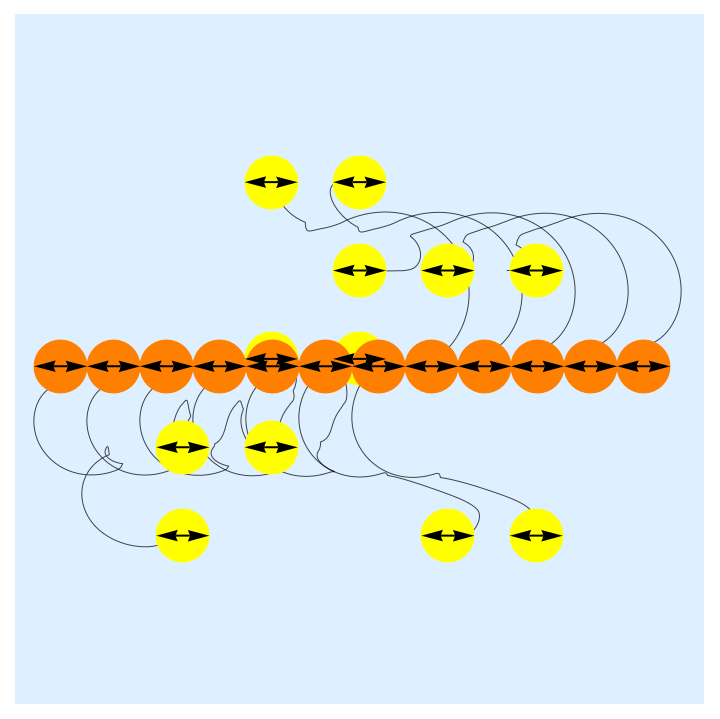}
    \caption{Pullers (far zone)}
  \end{subfigure}

  \vspace{0.8em}

  \begin{subfigure}[t]{0.35\textwidth}
    \centering
    \includegraphics[width=\linewidth]{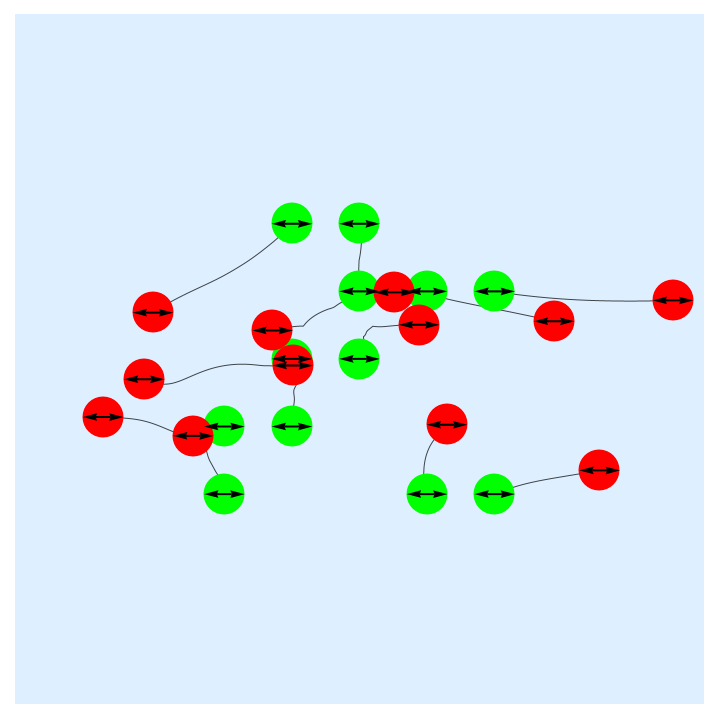}
    \caption{Pushers (near zone)}
  \end{subfigure}
  \hfill
  \begin{subfigure}[t]{0.35\textwidth}
    \centering
    \includegraphics[width=\linewidth]{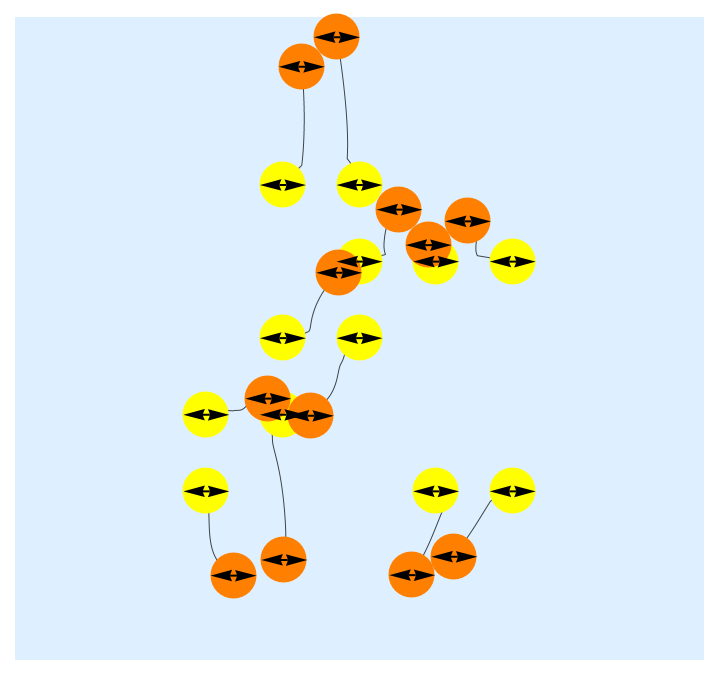}
    \caption{Pullers (near zone)}
  \end{subfigure}

  \caption{
Far- and near-zone dynamics of co-aligned clusters of twelve active particles in an incompressible membrane.
Columns correspond to pushers (left) and pullers (right).
Particles are initialized at random locations within a disc.
Initial positions are shown in green (pushers) and yellow (pullers),
final positions in red and orange, respectively, with trajectories shown in grey.
A soft harmonic repulsion is included to prevent particle overlap.
}
  \label{fig:fz-nz-combined}
\end{figure*}

\begin{figure}[t]
  \centering
  \includegraphics[width=0.55\linewidth]{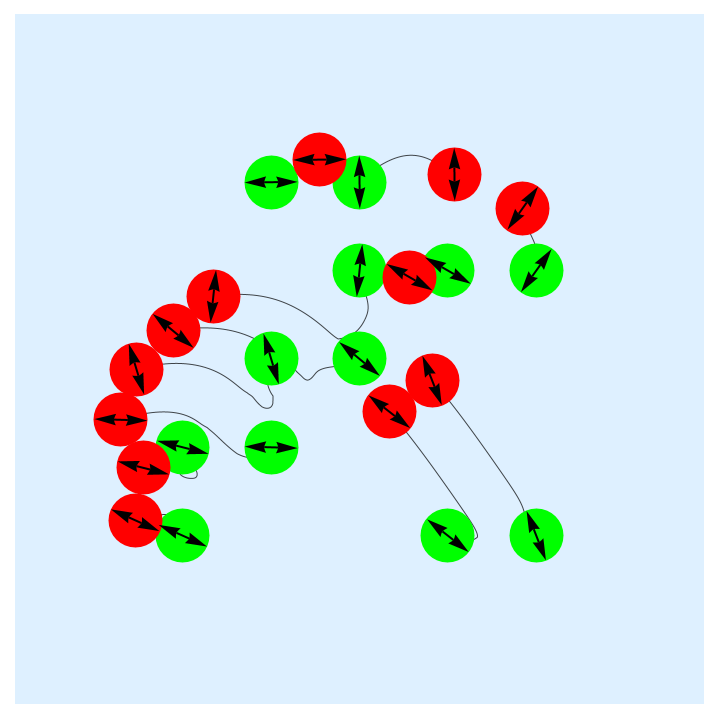}
 \caption{
Representative far–zone dynamics of a randomly initialized cluster of twelve pushers in the incompressible membrane with \emph{random} initial locations (within a disc) and \emph{random} orientations. Green disks indicate initial positions, red disks indicate final positions, and the grey curves trace the particle trajectories. A soft harmonic repulsion is included to regularize close encounters.}
  \label{repr}
\end{figure}

The regularization scheme used in
Appendices~\ref{app:numerics_near} and~\ref{app:numerics_far}
consists of a soft regulator introduced directly in the velocity kernels.
As a robustness check, the simulations in Appendix~\ref{supfig} instead
employ an explicit short-range repulsion. We introduce a harmonic contact
potential
\begin{equation}
V(r_{ij}) =
\begin{cases}
\dfrac{1}{2}k(\sigma - r_{ij})^2, & r_{ij} < \sigma, \\
0, & r_{ij} \ge \sigma,
\end{cases}
\end{equation}
with force $\mathbf F_{ij}=k(\sigma-r_{ij})\hat{\mathbf r}_{ij}$ for
$r_{ij}<\sigma$. The qualitative behaviour is unchanged: far-field
interactions produce aggregation, while near-field interactions do not,
demonstrating that the conclusions are independent of the regularization scheme.

\end{document}